\documentclass[twocolumn, tighten, numberedappendix]{aastex62}

\usepackage{mathrsfs}
\usepackage{savesym}
\savesymbol{tablenum}
\usepackage[separate-uncertainty=true, multi-part-units=single]{siunitx}
\DeclareSIUnit\jansky{Jy}
\restoresymbol{SIX}{tablenum}

\usepackage[normalem]{ulem}
\usepackage{lipsum}
\usepackage[utf8x]{inputenc}
\usepackage[encapsulated]{CJK}
\usepackage{ucs}
\newcommand{\cntext}[1]{\begin{CJK}{UTF8}{gbsn}#1\end{CJK}}
\usepackage{newtxtext}
\usepackage[varg]{newtxmath}
\usepackage{ulem}
\shortauthors{Appel et al.}
\usepackage{devanagari}
\usepackage{tablefootnote}

\newcommand{\medianDetNET}{258}

\newcommand{\fsky}{75}
\newcommand{\observingArea}{31000}
\newcommand{\centerfreq}{38.0}
\newcommand{\centerfreqps}{38.5}
\newcommand{\centerfreqerr}{0.2}
\newcommand{\ndet}{64}
\newcommand{\psat}{6.3}
\newcommand{\ptes}{4.2}
\newcommand{\popt}{1.6}
\newcommand{\Topt}{21}
\newcommand{\pd}{0.5}
\newcommand{\resp}{-8.2}
\newcommand{\tauopt}{3.4}
\newcommand{\tauC}{17}
\newcommand{\heatC}{3}
\newcommand{\thermalG}{177}
\newcommand{\thermalkappa}{13.4}
\newcommand{\Tc}{149}
\newcommand{\Rn}{8.2}
\newcommand{\Rshunt}{0.25}
\newcommand{\Tsys}{27}
\newcommand{\Tdet}{6}
\newcommand{\eff}{0.48}
\newcommand{\efferr}{0.02}
\newcommand{\dTrjdP}{13.1}
\newcommand{\dTrjdPerr}{0.3}
\newcommand{\dTcmbdTrj}{1.04}
\newcommand{\dTcmbdPerr}{0.3}
\newcommand{\dTcmbdP}{13.6}
\newcommand{\nepd}{11}
\newcommand{\nep}{19}
\newcommand{\arraynet}{32}
\newcommand{\bandwidth}{11.4}
\newcommand{\solidangle}{796}
\newcommand{\sderr}{7} 
\newcommand{\dilution}{0.077}
\newcommand{\pointflux}{27.6}
\newcommand{\pointfluxtauA}{27.8}
\newcommand{\pointfluxerr}{0.5}

\newcommand{\Tm}{210}
\newcommand{\poffset}{16.0} 
\newcommand{\pamp}{2.2} 
\newcommand{\phaseoffset}{3.3}
\newcommand{\moonperiod}{29.5}

\newcommand{\tauAT}{8.56}
\newcommand{\tauATerr}{0.27}
\newcommand{\tauAJy}{308}
\newcommand{\tauAJyerr}{11} 
\newcommand{\centerfreqtaua}{38.4}

\begin{document}

\title{On-sky performance of the CLASS Q-band telescope}


\author[0000-0002-8412-630X]{John W.~Appel}
\affiliation{Department of Physics and Astronomy, Johns Hopkins University, 
3701 San Martin Drive, Baltimore, MD 21218, USA}



\author[0000-0001-5112-2567]{Zhilei Xu (\cntext{徐智磊}\!)}
\affiliation{Department of Physics and Astronomy, University of Pennsylvania, 
209 South 33rd Street, Philadelphia, PA 19104, USA}
\affiliation{Department of Physics and Astronomy, Johns Hopkins University, 
3701 San Martin Drive, Baltimore, MD 21218, USA}

\author{Ivan L.~Padilla}
\affiliation{Department of Physics and Astronomy, Johns Hopkins University,
3701 San Martin Drive, Baltimore, MD 21218, USA}

\author[0000-0003-1248-9563]{Kathleen Harrington}
\affiliation{Department of Physics and Astronomy, Johns Hopkins University,
3701 San Martin Drive, Baltimore, MD 21218, USA}
\affiliation{Department of Physics, University of Michigan, Ann Arbor, MI, 48109, USA}

\author{Basti\'an Pradenas Marquez}
\affiliation{Departamento de F\'isica, FCFM, Universidad de Chile, Blanco Encalada 2008, Santiago, Chile}

\author{Aamir Ali}
\affiliation{Department of Physics,
University Of California,
Berkeley, CA 94720, USA}
\affiliation{Department of Physics and Astronomy, Johns Hopkins University,
3701 San Martin Drive, Baltimore, MD 21218, USA}

\author[0000-0001-8839-7206]{Charles L.~Bennett}
\affiliation{Department of Physics and Astronomy, Johns Hopkins University, 
3701 San Martin Drive, Baltimore, MD 21218, USA}

\author{Michael K. Brewer}
\affiliation{Department of Physics and Astronomy, Johns Hopkins University,
3701 San Martin Drive, Baltimore, MD 21218, USA}

\author[0000-0001-8468-9391]{Ricardo Bustos}
\affiliation{Facultad de Ingenier\'ia, Universidad Cat\'olica de la Sant\'isima Concepci\'on, Alonso de Ribera
2850, Concepci\'on, Chile}

\author[0000-0003-1127-0965]{Manwei Chan}
\affiliation{Department of Physics and Astronomy, Johns Hopkins University,
3701 San Martin Drive, Baltimore, MD 21218, USA}

\author[0000-0003-0016-0533]{David T.~Chuss}
\affiliation{Department of Physics, Villanova University, 800 Lancaster Avenue, Villanova, PA 19085, USA
}

\author{Joseph Cleary}
\affiliation{Department of Physics and Astronomy, Johns Hopkins University,
3701 San Martin Drive, Baltimore, MD 21218, USA}

\author{Jullianna Couto}
\affiliation{Department of Physics and Astronomy, Johns Hopkins University,
3701 San Martin Drive, Baltimore, MD 21218, USA}

\author[0000-0002-1708-5464]{Sumit Dahal ({\dn \7{s}Emt dAhAl})}
\affiliation{Department of Physics and Astronomy, Johns Hopkins University,
3701 San Martin Drive, Baltimore, MD 21218, USA}

\author{Kevin Denis}
\affiliation{Goddard Space Flight Center, 8800 Greenbelt Road, Greenbelt, MD 20771, USA}

\author{Rolando D\"unner}
\affiliation{Instituto de Astrof\'isica and Centro de Astro-Ingenier\'ia, Facultad de F\'isica, Pontificia Universidad Cat\'olica de Chile, Av. Vicu\~na Mackenna 4860, 7820436 Macul, Santiago, Chile}

\author[0000-0001-6976-180X]{Joseph R.~Eimer}
\affiliation{Department of Physics and Astronomy, Johns Hopkins University,
3701 San Martin Drive, Baltimore, MD 21218, USA}

\author[0000-0002-4782-3851]{Thomas~Essinger-Hileman}
\affiliation{Goddard Space Flight Center, 8800 Greenbelt Road, Greenbelt, MD 20771, USA}

\author[0000-0002-2061-0063]{Pedro Fluxa}
\affiliation{Instituto de Astrof\'isica and Centro de Astro-Ingenier\'ia, Facultad de F\'isica, Pontificia Universidad Cat\'olica de Chile, Av. Vicu\~na Mackenna 4860, 7820436 Macul, Santiago, Chile}

\author{Dominik Gothe}
\affiliation{Department of Physics and Astronomy, Johns Hopkins University,
3701 San Martin Drive, Baltimore, MD 21218, USA}

\author{Gene C. Hilton}
\affiliation{Quantum Sensors Group, National Institute of Standards and Technology, 325 Broadway, Boulder, CO 80305, USA}

\author{Johannes Hubmayr}
\affiliation{Quantum Sensors Group, National Institute of Standards and Technology, 325 Broadway, Boulder, CO 80305, USA}

\author{Jeffrey Iuliano}
\affiliation{Department of Physics and Astronomy, Johns Hopkins University,
3701 San Martin Drive, Baltimore, MD 21218, USA}

\author{John Karakla}
\affiliation{Department of Physics and Astronomy, Johns Hopkins University,
3701 San Martin Drive, Baltimore, MD 21218, USA}

\author[0000-0003-4496-6520]{Tobias A.~Marriage}
\affiliation{Department of Physics and Astronomy, Johns Hopkins University, 
3701 San Martin Drive, Baltimore, MD 21218, USA}

\author{Nathan J.~Miller}
\affiliation{Department of Physics and Astronomy, Johns Hopkins University, 
3701 San Martin Drive, Baltimore, MD 21218, USA}
\affiliation{Goddard Space Flight Center, 8800 Greenbelt Road, Greenbelt, MD 20771, USA}

\author{Carolina N\'u\~nez}
\affiliation{Department of Physics and Astronomy, Johns Hopkins University,
3701 San Martin Drive, Baltimore, MD 21218, USA}

\author{Lucas Parker}
\affiliation{Space and Remote Sensing, MS D436, Los Alamos National Laboratory,
Los Alamos, NM 87544, USA}
\affiliation{Department of Physics and Astronomy, Johns Hopkins University, 3701 San Martin Drive, Baltimore, MD 21218, USA}

\author[0000-0002-4436-4215]{Matthew Petroff}
\affiliation{Department of Physics and Astronomy, Johns Hopkins University,
3701 San Martin Drive, Baltimore, MD 21218, USA}

\author{Carl D. Reintsema}
\affiliation{Quantum Sensors Group, National Institute of Standards and Technology, 325 Broadway, Boulder, CO 80305, USA}

\author{Karwan Rostem}
\affiliation{Goddard Space Flight Center, 8800 Greenbelt Road, Greenbelt, MD 20771, USA}

\author{Robert W. Stevens}
\affiliation{Quantum Sensors Group, National Institute of Standards and Technology, 325 Broadway, Boulder, CO 80305, USA}

\author{Deniz Augusto Nunes Valle}
\affiliation{Department of Physics and Astronomy, Johns Hopkins University,
3701 San Martin Drive, Baltimore, MD 21218, USA}

\author[0000-0001-9269-5046]{Bingjie Wang (\cntext{王冰洁}\!)}
\affiliation{Department of Physics and Astronomy, Johns Hopkins University,
3701 San Martin Drive, Baltimore, MD 21218, USA}

\author[0000-0002-5437-6121]{Duncan J.~Watts}
\affiliation{Department of Physics and Astronomy, Johns Hopkins University, 3701 San Martin Drive, Baltimore, MD 21218, USA}

\author[0000-0002-7567-4451]{Edward J.~Wollack}
\affiliation{Goddard Space Flight Center, 8800 Greenbelt Road, Greenbelt, MD 20771, USA}

\author[0000-0001-6924-9072]{Lingzhen Zeng}
\affiliation{Harvard-Smithsonian Center for Astrophysics: Cambridge, MA, USA}

\correspondingauthor{John W.~Appel}
\email{jappel3@jhu.edu}

\published{\today}
\submitjournal{\apj}

\begin{abstract}
\added{ The Cosmology Large Angular Scale Surveyor (CLASS) is mapping the polarization of the cosmic microwave background (CMB) at large angular scales ($2<\ell\lesssim200$) in search of a primordial gravitational wave B-mode signal down to a tensor-to-scalar ratio of $r \approx 0.01$. The same dataset will provide a near sample-variance-limited measurement of the optical depth to reionization. Between June 2016 and March 2018, CLASS completed the largest ground-based Q-band CMB survey to date, covering over \num{\observingArea}~square-degrees (\fsky\% of the sky), with an instantaneous array noise-equivalent temperature (NET) sensitivity of \SI{\arraynet}{\micro\kelvin_{\mathrm{cmb}}\sqrt{\second}}.
We demonstrate that the detector optical loading (\SI{\popt}{\pico\watt}) and noise-equivalent power (\SI{\nep}{\atto\watt\sqrt{\second}}) match the expected noise model dominated by photon bunching noise.
We derive a \SI{\dTrjdP\pm\dTrjdPerr}{\kelvin\per\pico\watt} calibration to antenna temperature based on Moon observations, which translates to an optical efficiency of $\eff\pm\efferr$ and a \SI{\Tsys}{\kelvin} system noise temperature. 
Finally, we report a Tau~A flux density of \SI{\tauAJy\pm\tauAJyerr}{\jansky} at \SI[multi-part-units=single]{\centerfreqtaua\pm\centerfreqerr}{\giga\hertz}, consistent with the \textit{WMAP} Tau~A time-dependent spectral flux density model. 
}
\end{abstract}

\section{Introduction}\label{sec:intro}

Mapping the polarization of the cosmic microwave background (CMB) is essential for understanding the earliest moments of the Universe. In addition to constraining inflation~\citep{guth:1981, sato:1981,linde:1982,starobinsky:1982,albrecht/steinhardt:1982, planck_2018_inflation} and the standard six-parameter $\Lambda$CDM model~\citep{Hinshaw2013,planck_2018_cosmo_param},  the polarization of the CMB is a probe for the epoch of reionization and the growth of large-scale structure. 
 The \SI{100}{\micro\kelvin} CMB intensity fluctuations are polarized by Thomson scattering at the few percent level \citep{rees_cmbpol,dasi}. This polarization is decomposed into E modes, which provide our best constraint on the optical depth to reionization~\citep{Hinshaw2013,planck_2018_cosmo_param}, and B modes, which probe inflationary gravitational radiation~\citep{kamionkowski,seljak_zaldariaga}. The B-mode component is at least ten times fainter than the E-mode component~\citep{bicep_keck_2018}.
Both must be separated from polarized Galactic emission (e.g.~\cite{bicep2016}). Averaged over the sky at high galactic latitudes, polarized  dust emission
is the  dominant Galactic component at frequencies above \SI{70}{\giga\hertz}~\citep{planck:IntermediateXIX,planck2015XX,planckL2016,planck_2018_dust}, while synchrotron 
is the strongest polarized emission mechanism at lower frequencies~\citep{planck_2018_diffuse,bennett:2013}.
 On small angular scales gravitational lensing of E-modes induces a B-mode signal larger than the current upper limit on primordial inflationary B-modes. 
 Efforts toward characterizing these small angular scale B-modes include \cite{polarbear_2014}, \cite{spt_2018}, and \cite{actpol_2017}.

The Cosmology Large Angular Scale Surveyor (CLASS) will measure the polarized microwave sky in bands centered at approximately \SIlist{40;90;150;220}{\giga\hertz} from an altitude of \SI{5200}{\meter} above sea level in the Atacama Desert of northern Chile~\citep{tom_spie,harrington2016} inside the Parque Astron\'omico Atacama~\citep{parque_atacama}. 
The Q-band (\SI{40}{\giga\hertz}) telescope probes synchrotron emission~\citep{joseph_SPIE,spie_jappel}, whereas the G-band (dichroic \SIlist{150;220}{\giga\hertz}) telescope maps dust.
Two W-band (\SI{90}{\giga\hertz}) telescopes provide the necessary sensitivity to the CMB polarized signal~\citep{sumit_spie_2018}. 
The location, design, and survey strategy of the CLASS telescopes are defined to reconstruct the microwave polarization at large angular scales (multipoles $2<\ell\lesssim200$) over 75\% of the sky.
To achieve this goal, CLASS employs a Variable-delay Polarization Modulator (VPM) as its first optical element to increase stability and mitigate instrumental polarization~\citep{Miller2015,chuss_vpm_2012}.
The benefits of implementing a fast ($\sim$\SI{10}{\hertz}) polarization modulator as your first optical element has been demonstrated from the ground by the Atacama B-mode Search experiment~\citep{abs_hwp, abs_final}.
Telescope boresight rotation and a comoving ground shield mitigate contamination by terrestrial polarization sources. 
The CLASS survey is forecast to constrain the optical depth to reionization $\tau$ to near the cosmic variance limit and the inflationary tensor-to-scalar ratio to $r\approx 0.01$~\citep{Watts2015,Watts2018}.
The optical depth is the least constrained $\Lambda$CDM parameter, and new measurements at $\ell<12$ are important for realizing the full potential of cosmological probes of neutrino masses \citep{Allison2015,Watts2018}.

\begin{figure}
\begin{center}
\includegraphics[trim={6cm 0 4cm 0},clip,width =0.25\textwidth]{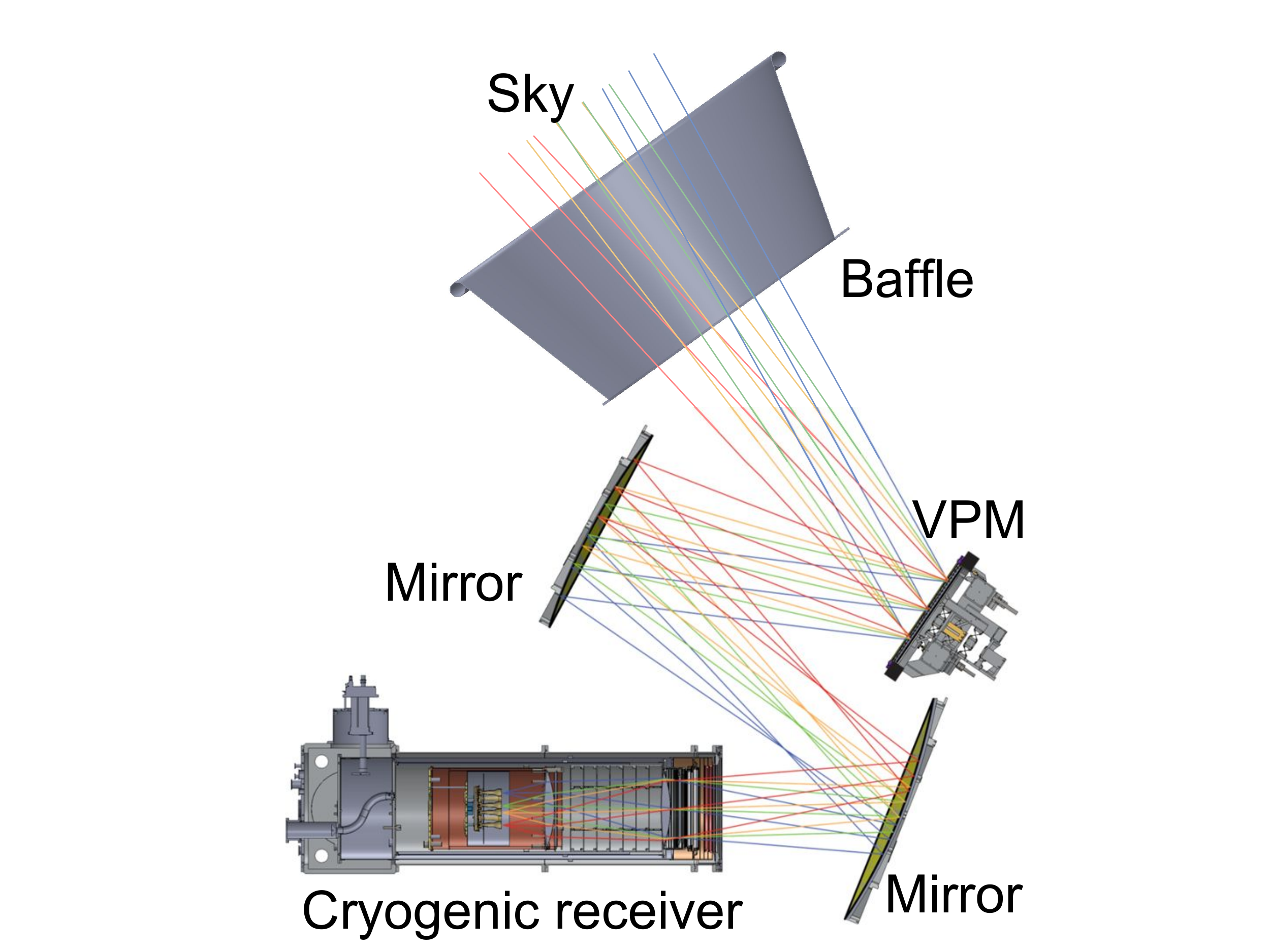}
\includegraphics[trim={0cm 0cm 0cm 0.3cm},clip,width =0.19\textwidth]{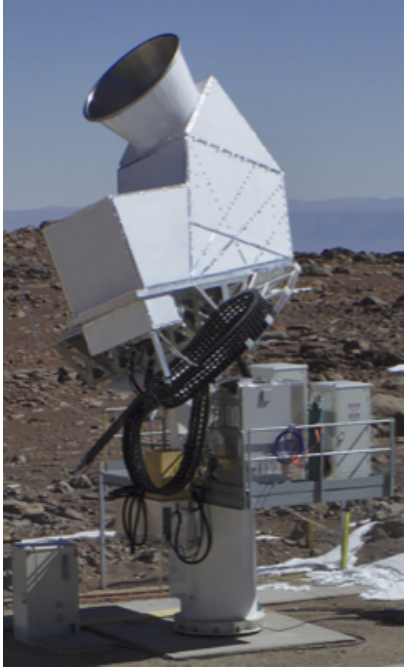}
\caption{\textit{Left:} Diagram of the Q-band instrument including baffle, VPM, mirrors, and cryostat ~\citep{joseph_SPIE,tom_spie,spie_jappel,harrington2016,Harrington2018}. For a closer view of the Q-band detector assembly and cryogenic receiver see: Figure 1 in \citet{karwan_spie_2014, spie_jappel}, and Figure 8 in \citet{harrington2016}. \textit{Right:} Photograph of the Q-band telescope configuration during the June 2016 to March 2018 observing period.
The telescope is enclosed in a metal structure that protects the instruments and prevents pickup from terrestrial sources.}
\label{fig:qband_diagram_pic}
\end{center}
\end{figure}

This is the first paper describing the on-sky performance of CLASS.
This analysis is based on observations with the CLASS Q-band telescope between June 2016 and March 2018 (see Figure~\ref{fig:qband_diagram_pic}).
In this paper we discuss the calibration and performance of the Q-band telescope for intensity measurements, leaving discussions of polarized performance to future papers.
In Section~\ref{sec:dark}, we present  median detector parameters extracted from $I$-$V$ measurements and array sensitivity estimates based on the power spectral density (PSD) of the time-ordered data (TOD).
In Section~\ref{sec:moon_calib}, observations of the Moon are used to constrain the average detector beam, the relative gain between detectors, the telescope optical efficiency, and the calibration factor to convert power measured at the detectors to antenna temperature.
Using this Moon-based antenna temperature calibration, we present a new measurement of Tau~A flux density at Q band in Section~\ref{sec:tauA}.
Finally, Section~\ref{sec:discussion} summarizes the CLASS Q-band detector array performance during its first observing campaign.

\section{On-sky detector characteristics}\label{sec:dark}
\begin{table*}
\footnotesize
    \center
    \renewcommand{\arraystretch}{1.2}
    \begin{tabular}{p{4.4cm}p{2.2cm}}
    \multicolumn{2}{c}{TES Bolometer Parameters} \\
    \hline
    \hline
    Phonon~Power~($P_{\upvarphi}$) & \SI{\psat}{\pico\watt} \\
    Bias Power ($P$) & \SI{\ptes}{\pico\watt} \\
    Dark~Power~Offset~($P_{\mathrm{D}}$) & \SI{\pd}{\pico\watt}\\
    Optical~Loading~($P_{\gamma}$) & \SI{\popt}{\pico\watt} \\ 
    Responsivity~($S$) & \hspace{-7.0pt}\SI{\resp}{\micro\ampere\per\pico\watt} \\
    Optical~Time~Constant~($\tau_{\gamma}$) & \SI{\tauopt}{\milli\second} \\
    Thermal~Time~Constant~($\tau_{\upvarphi}$) & \SI{\tauC}{\milli\second} \\
    Heat Capacity ($C$) & \SI{\heatC}{\pico\joule\per\kelvin}\\
    Thermal Conductivity ($G$) & \SI{\thermalG}{\pico\watt\per\kelvin} \\
    Thermal Conductivity Constant ($\kappa$) & \SI{\thermalkappa}{\nano\watt~\kelvin^{-4}} \\
    Critical Temperature ($T_{\mathrm{c}}$) &
    \SI{\Tc}{\milli\kelvin} \\
    Normal Resistance  ($R_\mathrm{N}$) & \SI{\Rn}{\milli\ohm} \\
    Shunt Resistance  ($R_\mathrm{sh}$) & \SI{\Rshunt}{\milli\ohm} \\
    TES loop Inductance ($L$)& \SI{500}{\nano\henry} \\ 
    \hline
    \hline
    \end{tabular}
    \quad
    \begin{tabular}{p{4.4cm}p{2.2cm}}
    \multicolumn{2}{c}{Optical Performance Parameters} \\
    \hline
    \hline
    System~Noise~Temperature~(${T}_{\mathrm{sys}}$) & \SI{\Tsys}{\kelvin} \\
    Telescope~Efficiency~($\eta$) & \eff \\
    RJ Temp Calibration~($\frac{dT_{\mathrm{RJ}}}{dP_{\gamma}}$) &  \SI{\dTrjdP}{\kelvin\per\pico\watt}  \\
    CMB-RJ Calibration ($\frac{dT_{\mathrm{cmb}}}{dT_{\mathrm{RJ}}}$) &  \dTcmbdTrj  \\
    Detector~Dark~Noise~Power~($\mathrm{NEP}_\mathrm{d}$) & \SI{\nepd}{\atto\watt\sqrt{\second}} \\
    Detector Total~Noise~Power~($\mathrm{NEP}$) & \SI{\nep}{\atto\watt\sqrt{\second}} \\
    Detector Noise Temperature ($\mathrm{NET}$) & \SI{\medianDetNET}{\micro\kelvin_{\mathrm{cmb}}\sqrt{\second}}\\
    Optical Detectors ($N_{\mathrm{det}}$) & \ndet \\
    Array Noise Temperature & \SI{\arraynet}{\micro\kelvin_{\mathrm{cmb}}\sqrt{\second}} \\
    RJ Extended Source Band Center ($\nu_\mathrm{o}$) & \SI{\centerfreq}{\giga\hertz} \\
    RJ Point Source Band Center ($\nu'_\mathrm{o}$) & \SI{\centerfreqps}{\giga\hertz} \\
    Bandwidth ($\Delta \nu$) &  \SI{\bandwidth}{\giga\hertz} \\
    Beam Solid Angle ($\Omega$) & \SI{\solidangle}{\micro\steradian} \\
    RJ Point Source Flux Factor ($\Gamma$) & \SI{\pointflux}{\micro\kelvin\per\jansky}\\
    \hline
    \hline
    \end{tabular}
    \renewcommand{\arraystretch}{1}
\caption{Table of Q-band detector parameters. Parameters on the left column are derived with the help of $I$-$V$ data and represent the median value across the detector array. Time constant and heat capacity estimates also depend on measurements of the VPM synchronous signal.
The right-hand side parameters are derived using: TOD power spectral density near the 10~Hz modulation frequency to estimate NEP, Moon observations to measure the beam solid angle and calibrate power at the bolometer to antenna temperature, and laboratory Fourier-Transform Spectrometer (FTS) measurements to determine bandpass properties. }
\label{table:onsky_med}
\end{table*}

The Q-band array consists of 36 feedhorn-coupled, dual-polarization detectors. Each polarimeter has two transition edge sensor (TES) bolometers, one for measuring the optical power in each orthogonal linear polarization channel \citep{spie_jappel,karwan_spie,dchuss_qdet_2012,dchuss_qdet_2014}.
The bolometers are read out through time-division multiplexing (TDM) of superconducting quantum interference device (SQUID) amplifiers \citep{nist_tdm_mux13b,ubc_mce}. 
The CLASS Q-band two-stage TDM scheme consists of \num{8} columns multiplexing \num{11} rows of SQUIDs, for a total of \num{88} channels, of which \num{14} are dedicated dark SQUID channels used to characterize readout noise and magnetic field pickup. 
A dark SQUID is a readout channel that is not connected to a TES bolometer.
Two readout channels are connected to dark TES bolometers fabricated within the polarimeter chips \citep{denis_fabq,denis_fabw}. 
Unlike the optical bolometers, the dark bolometers are not connected to antennas at the waveguide output of the feedhorns \citep{dchuss_qdet_2012,wollack_omt}.
Of the \num{72} polarization sensitive bolometers, \ndet~were operational during the first observing campaign; the remaining eight channels were lost during deployment due to a readout electronics failure. These channels were recovered for the second observing campaign, which began in June 2018. 

The Q-band telescope observed on a 24-hour cycle that started routinely at 14:00 UTC (late-morning local time).
The Q-band receiver operates a dilution refrigerator that continuously cools the detector array to $T_{\mathrm{b}} \approx \SI{42}{\milli\kelvin}$~\citep{jeff_spie_2018} during science operations, therefore allowing any observation cadence.
Our 24 hour cycle is chosen to yield a full sky map each day at one boresight. We change boresight angle everyday, and the timing of the schedule end/start coincides with the site crew work schedule.
At the beginning and end of an observation cycle,
the detector bias voltage ($V$) was swept while recording the current response ($I$) to produce what will hereafter be called an ``$I$-$V$ curve.''
Additional $I$-$V$ curves are acquired  before special data sets such as wire-grid calibration measurements, and detector noise tests with the cryostat window covered.
These $I$-$V$ curves are used to choose the optimal bias voltage for each column composed of up to 10 TES bolometers.
During observations these bias voltages place the array TES bolometers on their superconducting transition between 30\% and 60\% of their normal resistance.
The detector saturation power ($P_{\mathrm{sat}}$) is extracted from $I$-$V$ data and defined as the detector bias power ($P = IV$) evaluated at 80\% of the TES normal resistance ($R_\mathrm{N}$).
The difference between the $P_{\mathrm{sat}}$ measured in dark laboratory tests with the detectors enclosed in a \SI{1}{\kelvin} cavity ($P_{\upvarphi}$), and those measured while observing the sky is interpreted as the optical power loading ($P_{\gamma}$) on the detectors.
Included in this number is a correction for a small offset tracked by neighboring nonoptical bolometers ($P_{\mathrm{D}}$) discussed in section~\ref{sec:opt_load}.

Detector responsivity ($S = \mathrm{d}I/\mathrm{d}P_{\gamma}$) estimated from $I$-$V$ data is used to calibrate current signals ($\mathrm{d}I$) across the TES to power deposited on the bolometer ($\mathrm{d}P_{\gamma}$).  
Detector optical time constants are extracted from the delayed response to the VPM synchronous signal (see Figure~\ref{fig:beam_fts_tau}) that appears at the modulation frequency of \SI{10}{\hertz}.
Combining measured time constants with $I$-$V$ curve data, we derive the heat capacity of the bolometers.
Table \ref{table:onsky_med} summarizes median detector parameters across the array during this period.

\subsection{Optical loading}
\label{sec:opt_load}

\begin{figure*}
\begin{center}
\includegraphics[trim={0cm 0 0cm 0},clip,width =1.0\textwidth]{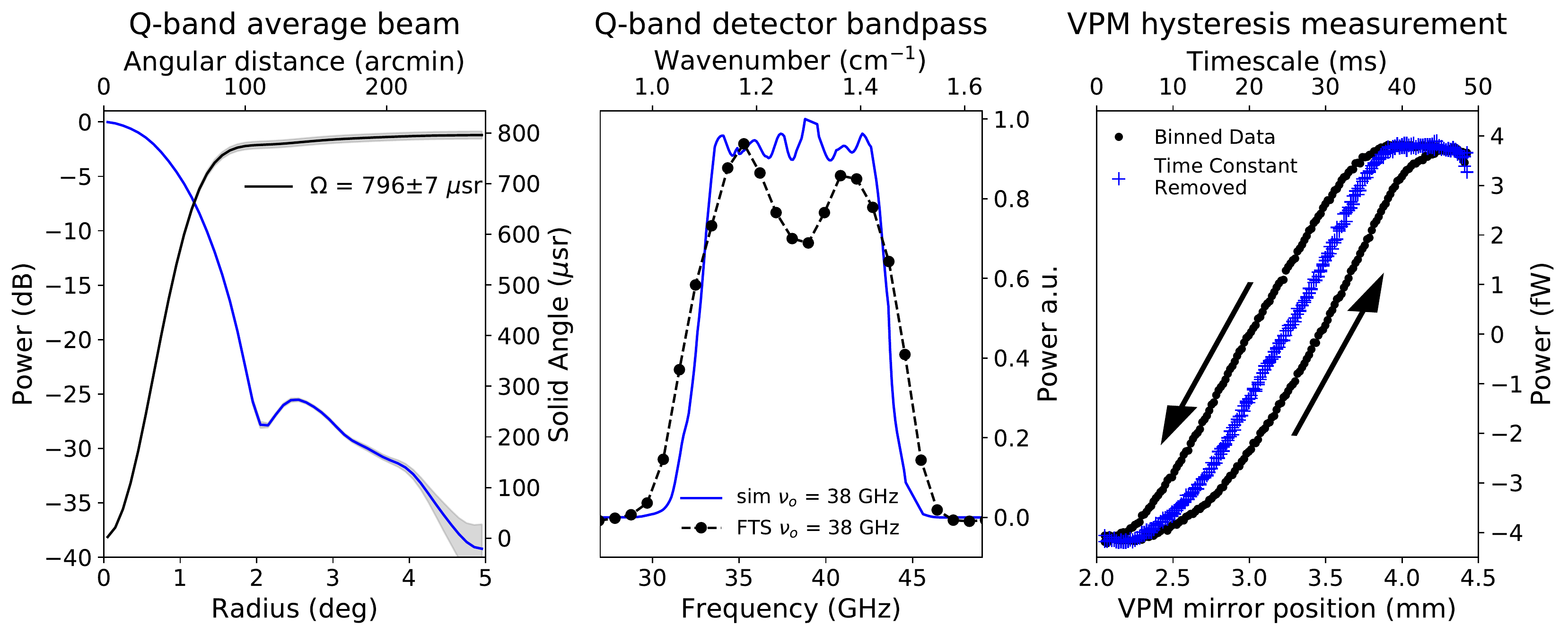}
\caption{\textit{Left:} The blue solid line shows the array-averaged radial beam profile derived from Moon observations, and the gray shaded region the measurement uncertainty. The main lobe is similar to a 1.5~degree Gaussian beam. The black line shows the array-averaged beam solid angle versus radius with a total solid angle of \SI{\solidangle}{\micro\steradian} (Xu et al., in prep.). \textit{Center:} The solid line plots the simulated detector bandpass~\citep{dchuss_qdet_2012} with center frequency \SI{38.0}{\giga\hertz} and bandwidth \SI{\bandwidth}{\giga\hertz}. 
The connected dots show a Q-band detector bandpass extracted from fourier transform spectrometer (FTS) measurements made in the laboratory (see \citet{sumit_spie_2018} for a description of the FTS testing setup).
We use a Martin-Puplett FTS~\citep[T. Wei 2010, private communication;][]{mp_fts} with a liquid nitrogen cooled blackbody~\citep{petroff_bb} as the input source while the FTS output is directed at a CLASS feedhorn coupled Q-band bolometer placed one meter behind a ten centimeter diameter cold stop.
The feedhorn's frequency dependent gain~\citep{zeng_phd} and the transmission of the lab-cryostat filters are divided out from the raw FTS measurement.  
The center frequency of the measured bandpass matches the simulation and has an estimated uncertainty of \SI{\centerfreqerr}{\giga\hertz} driven by the lab-cryostat filter transmission model.
The $\sim\SI{1}{\giga\hertz}$ resolution of the FTS broadens the measured bandpass edges compared to the simulation.
The remaining differences between the measured and simulated bandpass are likely due to unaccounted systematics of the FTS and/or the coupling optics.
The detector bandpass corresponds to the effective telescope bandpass for a beam filling extended Rayleigh-Jeans (RJ) thermal source. 
The effective center frequency for a Rayleigh-Jeans point source is \SI{\centerfreqps}{\giga\hertz}, obtained from combining the detector bandpass with the telescope frequency dependent gain (Fluxa, in prep.).
\textit{Right:} The raw (black) and time-constant corrected (blue) detector response to the VPM synchronous signal binned with respect to the grid-mirror distance (Harrington et al., in prep.). 
The black arrows indicate the direction of the raw signal as the VPM mirror is moved.}
\label{fig:beam_fts_tau}
\end{center}
\end{figure*}

The in-band (see bandpass in Figure~\ref{fig:beam_fts_tau}) optical power $P_{\gamma}$ dissipated on each bolometer is equal to the difference between the on-sky detector bias power $P$ and the phonon power $P_{\upvarphi}$ that flows from the bolometer island to the bath:
\begin{equation}
P_{\gamma} = P_{\upvarphi}-P,
\label{p_eq2}
\end{equation}
where $P_{\upvarphi}$  and the bias power $P$ are both measured with the detector baseplate temperature at $T_\mathrm{b}\approx\SI{50}{\milli\kelvin}$
($P_{\upvarphi}$ is equivalent to $P_{\mathrm{sat}}$ measured in dark laboratory tests with no optical loading).
The detector copper baseplate serves as both mechanical support and thermal heatsink for the detector chips \citep{spie_jappel}. 
Its temperature is tracked by a calibrated ruthenium oxide (ROX) temperature sensor.\footnote{RX-102A; https://www.lakeshore.com}

The Q-band array contains two dark TES bolometers that have similar electro-thermal properties to the optical detectors in the array, but are disconnected from the on-chip planar microwave circuitry that couples the radiation from a feedhorn to the optical TES bolometers in a pixel.
The saturation power for these bolometers decreases on average by $P_{\mathrm{D}}= \pd~$pW when opening the \SI{1}{\kelvin} detector cryostat volume to the sky.
Daily changes in atmospheric conditions affect both $P$ for optical detectors as well as $P_{\mathrm{D}}$.
In particular, we find that averaged across the observing period $P_{\mathrm{D}}/(P_{\upvarphi}-P) = 0.26$. 
The dark detector response to scanning an unresolved source like the Moon is $<$0.03\% that of the average optical detector. 
Hence the change in dark detector saturation power can be interpreted as an offset in $T_{\mathrm{b}}$ between the ROX and the silicon frame holding the bolometers, as opposed to optical coupling. 
This offset can be driven by changes in the \SI{1}{\kelvin} nylon filter~\citep{tom_spie,jeff_spie_2018} temperature ($\sim$\SI{4}{\kelvin} at its center) as atmospheric conditions change, since radiation emitted by the filter would fill the focal plane volume and weakly couple to the detector chip and/or the ROX.
We find less likely the alternative explanation of out-of-band power coupling directly to the TES island due to careful detector design isolating the TES bolometers from possible light leaks, and the lack of any out-of-band signal in our FTS measurements.
We assume dark detector saturation power offset is similar for all bolometers in the array; hence we subtract $P_{\mathrm{D}}$ from $P_{\gamma}$ for all optical channels.

The median in-band optical loading $P_{\gamma}= \SI{\popt}{\pico\watt}$ is consistent with the model presented in \citet{tom_spie} and \citet{spie_jappel}. 
Two factors deviate from the model: (1) slightly lower optical efficiency in the field reduces $P_{\gamma}$, and (2) the instrument's baffle and mount enclosure structure source $0.2~$pW of additional optical power.

\subsection{Detector responsivities and time constants}

 Ninety-eight percent of $I$-$V$ derived detector responsivities across all CMB observations fall between \SIlist{5;13}{\micro\ampere\per\pico\watt}.
This well-defined range is due to stable atmospheric loading, stable cryogenic temperatures, and near-optimal detector saturation powers~\citep{leg_precision}.

The VPM consists of a wire-grid that is placed in front of a mirror. The millimeter spacing between the mirror and the grid is optimized for the Q-band telescope \citep{Harrington2018}. In the CLASS telescopes, the mirror position is modulated at a frequency of \SI{10}{\hertz} to achieve polarization modulation.
In addition to reflecting and modulating the polarized sky signal, the VPM emits a small signal synchronous with the grid-mirror distance.
Each subset of CMB time-ordered data is fitted for a detector time constant that minimizes the hysteresis of this synchronous signal sourced by the VPM (see Figure \ref{fig:beam_fts_tau}). 
Eighty-six percent of CMB scans yield detector time constant ($\tau_{\gamma}$) measurements between \SIlist{2;6}{\milli\second}.
All are short enough to respond to the targeted 10~Hz modulation frequency and several of its harmonics.
Multiplying $\tau_{\gamma}$ by the electro-thermal feedback~\citep{irwin_hilton} speed-up factor estimated from $I$-$V$ data yields the detector thermal time constant ($\tau_{\upvarphi}$).  
The heat capacity ($C$) of each detector is then obtained by multiplying its average thermal time constant by the detector thermal conductivity ($G$).
The measured average bolometer heat capacity is \SI{\heatC}{\pico\joule\per\kelvin}.
All detector heat capacities are within \SI{1}{\pico\joule\per\kelvin} of the mean. 
Achieving the targeted heat capacity allows for stable/optimal biasing of the detectors in the field, improving detector sensitivity and observing efficiency.

\subsection{Detector Noise}

Detector noise performance is quantified in terms of noise equivalent power (NEP) at the bolometer.  We measure the NEP by averaging the power spectral density of the detector output in the side bands of the 10~Hz modulation frequency (9\textendash11~Hz).
To reduce correlated noise and improve the white noise estimate, we calculate individual detector NEP by first subtracting the TOD of detector pairs within a pixel (coupled to a feedhorn), then computing the power spectral density of the pair difference TOD and dividing by a factor of two in power squared units.
Here we do not consider single detectors whose pair is not operational; hence this NEP analysis focuses on 27 detector pairs.

The median single detector NEP in the first observing season is $\mathrm{NEP}=\SI{\nep}{\atto\watt\sqrt{\second}}$.
This result is consistent with expectations once we correct the design estimates in \cite{tom_spie} to account for photon bunching noise cross-terms, lower achieved optical efficiency, and additional beam spill onto the baffle and telescope enclosure structure.
Optical loading on the bolometers varies with atmospheric conditions; this allows us to probe the detector $\mathrm{NEP}$ vs. $P_{\gamma}$ relationship. 
Each $I$-$V$ measurement yields a $P_{\gamma}$ estimate for each detector, which corresponds to the $\mathrm{NEP}$ measured in the subsequent time-ordered data acquisition. We find that the change in $P_{\gamma}$ and $\mathrm{NEP}$ between consecutive $I$-$V$ measurements is small.

The NEP of a bolometer observing blackbody radiation is subject to both dark detector noise ($\mathrm{NEP}_\mathrm{d}$) and photon noise ($\mathrm{NEP}_\gamma$).
For the CLASS TES bolometers, $\mathrm{NEP}_\mathrm{d}$ is dominated by phonon thermal fluctuations but also contains contributions from TES Johnson noise and SQUID readout noise. Tests in dark laboratory cryostats yield an average Q-band array $\mathrm{NEP}_\mathrm{d} = \SI{\nepd}{\atto\watt\sqrt{\second}}$ \citep{spie_jappel}. 

The statistical properties of the photons emitted by thermal sources we observe (atmosphere, CMB, dielectric filters, Moon, etc.) generate noise fluctuations at the detector output, which cannot be suppressed by improving the detector characteristics \citep{richards_bolometer,zmuidzinas,vanvliet,mather_bolometer}.
The average variance in the number of photons ($n$) per mode sourced by a blackbody at temperature $T$ is $\langle(\Delta n)^2\rangle = \langle n \rangle + \langle n \rangle^2 $. 
The first term indicates the blackbody photons obey Poisson statistics in the limit that $n~\ll1$ ($kT/h\nu\ll1$), while in the limit $n~\gg1$ ($kT/h\nu\gg1$), the second term dominates and the photons arrive in bunches.
Here $k$ is Boltzmann's constant and $h$ is Planck's constant.
Photon counting statistics are translated to NEP ($\mathrm{NEP}_{\gamma}$) by identifying the spectral power density observed through a single mode detector as $P_\nu =  h\nu\langle n\rangle$; therefore~\citep{richards_bolometer}:
\begin{equation}    
(\mathrm{NEP}_{\gamma})^2 = \int h \nu P_\nu d\nu + \int P_\nu^2 d\nu \quad[\SI{}{\watt\squared\second}],
\label{eqn:nep_gamma}
\end{equation}
 where $ \int  P_\nu d\nu = P_{\gamma} \approx P_{\nu_\mathrm{o}} \Delta \nu \approx \eta k T \Delta \nu$ (since the CMB and other sources CLASS Q-band observes are close to the Rayleigh-Jeans limit), $\Delta \nu$ is the microwave signal bandwidth, and $\eta$ is the optical efficiency of the entire telescope system, including attenuation, reflection, and beam spill due to the detector, filters, lenses, window, mirrors, VPM, and baffle. 
 
\begin{figure}
\center
\includegraphics[trim={0cm 0cm 0cm 0},clip,width = 0.47\textwidth]{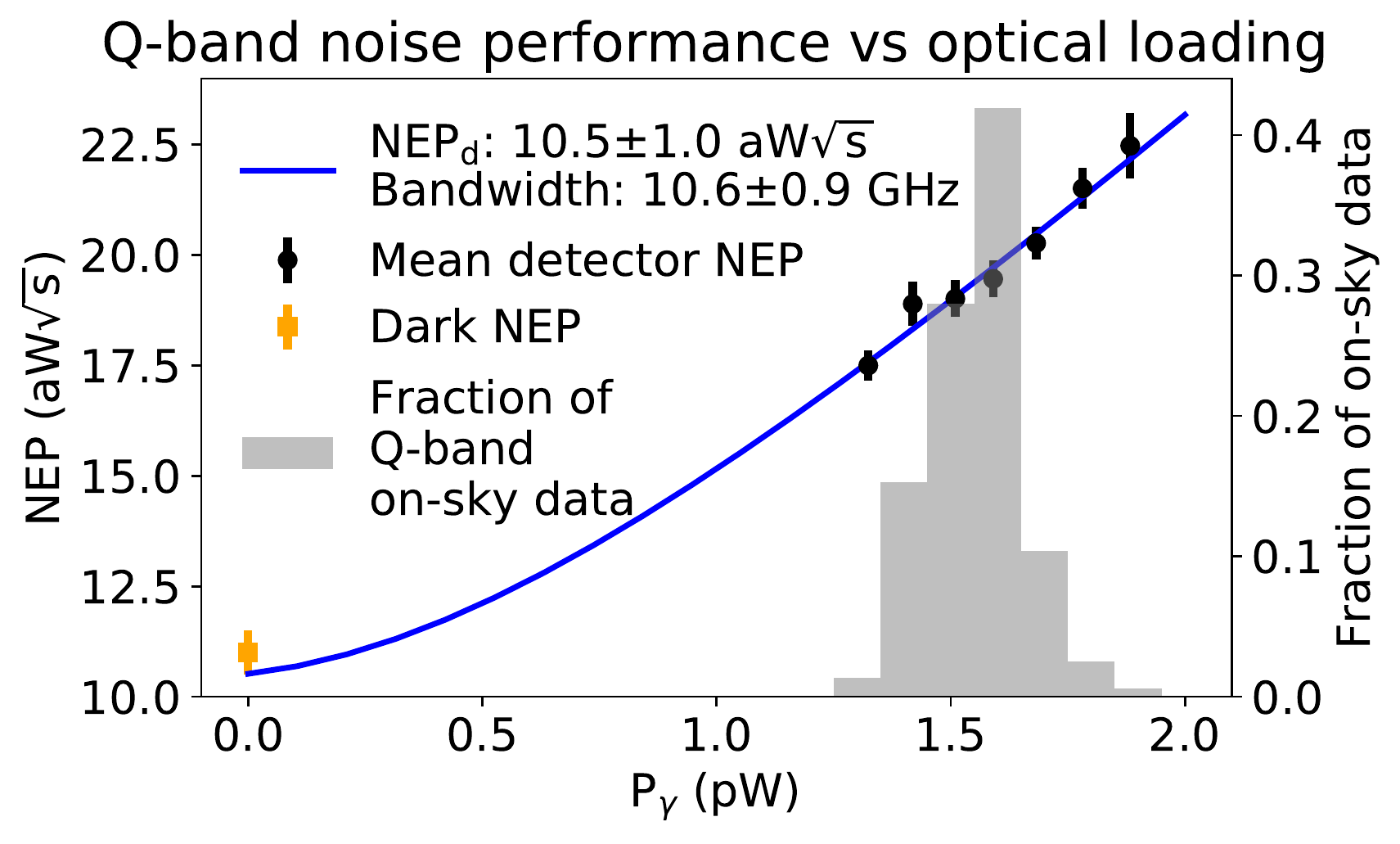}
\caption{The histogram in the figure shows the Q-band optical loading distribution during observations. The relatively narrow range highlights the stability of the atmosphere at the CLASS site in the Atacama Desert of Chile. 
The black data points are the average NEP across the detector array for each bin of optical loading.
The line is a model of on-sky NEP based on equation~\ref{eqn:nep} with the center frequency set to $\nu_\mathrm{o}= \SI{\centerfreq}{\giga\hertz}$, and the bandwidth $\Delta\nu$ and dark detector $\mathrm{NEP}_\mathrm{d}$ left as free parameters to fit. 
The results of the fit are consistent with independent FTS and dark laboratory measurements of $\Delta\nu$ and $\mathrm{NEP}_\mathrm{d}$ (see Table~\ref{table:onsky_med}). 
The measured $\mathrm{NEP}_\mathrm{d}$ is plotted in orange at zero optical loading.
Q-band $\mathrm{NEP}$ is dominated by bunching noise; therefore, sensitivity is driven by total optical loading and detector bandwidth.
The model provides a quantitative prediction of instrument sensitivity as optical loading changes with the telescope design. }
\label{fig:nep}
\end{figure}

 The total detector $\mathrm{NEP}$ can be expressed in terms of measured quantities $\mathrm{NEP}_{\mathrm{d}}$, $P_{\gamma}$,  $\Delta \nu$, and detector band center frequency $\nu_\mathrm{o}$ as:
 \begin{equation}    
(\mathrm{NEP})^2 =(\mathrm{NEP}_\mathrm{d})^2 + h \nu_\mathrm{o} P_\gamma +  \frac{P_\gamma^2} {\Delta\nu}  \quad[\SI{}{\watt\squared\second}].
\label{eqn:nep}
\end{equation}
Figure \ref{fig:nep} shows the measured $\mathrm{NEP}$ on the $y$-axis, and on the $x$-axis the corresponding measured $P_{\gamma}$. Equation \ref{eqn:nep} is fitted to the data points by setting $\nu_\mathrm{o} =$ \SI{\centerfreq}{\giga\hertz} and leaving $\mathrm{NEP}_{\mathrm{d}}$ and $\Delta \nu$ as free parameters. 
The best fit result of $\mathrm{NEP}_{\mathrm{d}} = \SI{10.5\pm1.0}{\atto\watt\sqrt{\second}}$ and $\Delta \nu = \SI{10.6\pm0.9}{\giga\hertz}$ is consistent with independent measurements of \SI{\nepd}{\atto\watt\sqrt{\second}} and \SI{\bandwidth}{\giga\hertz}. This confirms that the CLASS Q-band detectors are photon noise limited and that the NEP is dominated by the photon noise bunching term.
This NEP model provides a quantitative understanding of possible sensitivity improvements to the instrument if optical loading can be reduced without decreasing optical efficiency. In particular, \SI{300}{\kelvin} baffling configurations will be explored in future seasons, where control of systematic effects due to beam spill can be traded for sensitivity.

\section{Moon observations and calibration to antenna temperature}
\label{sec:moon_calib}

\begin{figure*}
\center
\includegraphics[width=\textwidth]{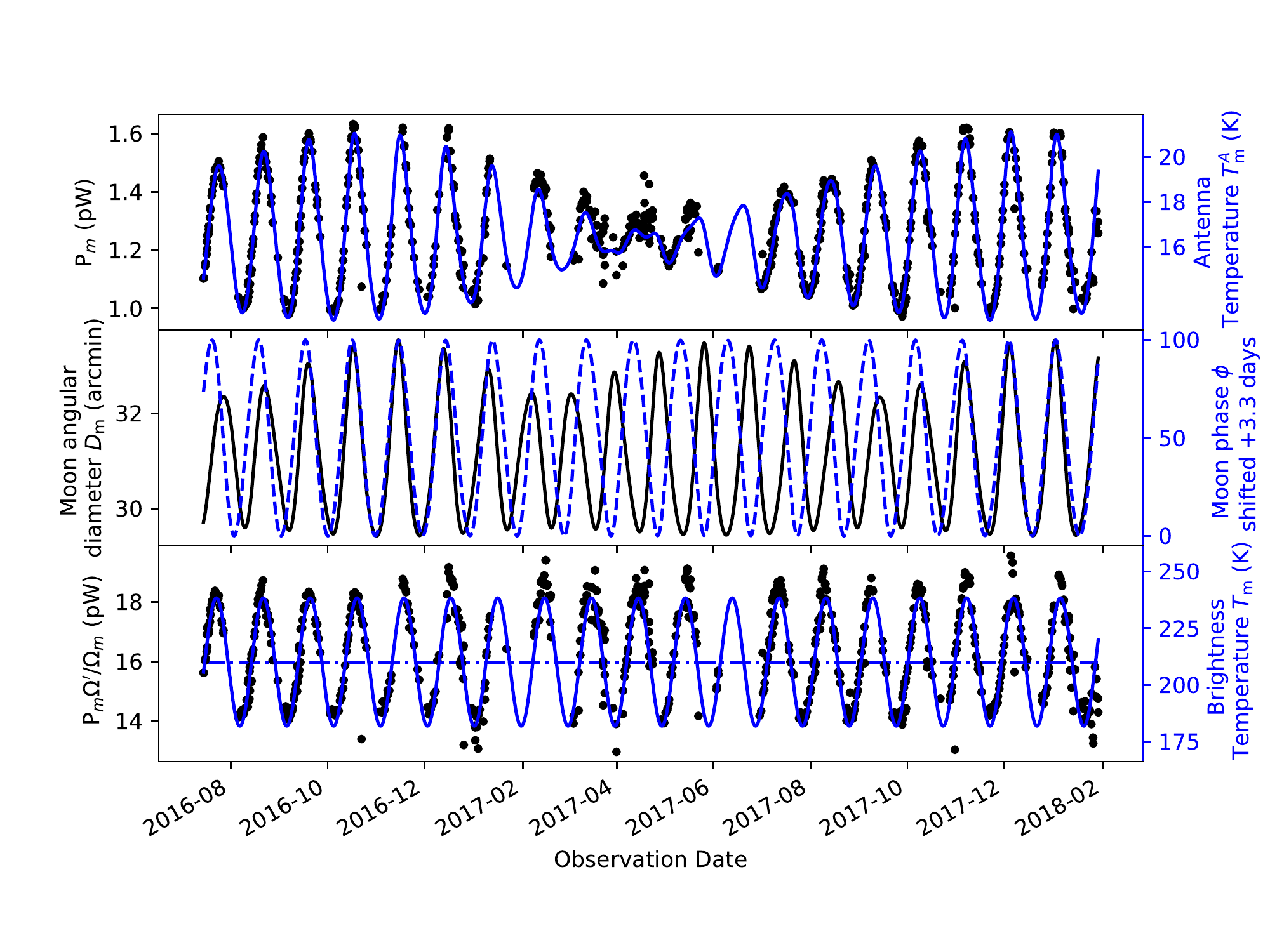}
\caption{ \textit{Top}: Data points are the array-averaged peak Moon power measured at the bolometers ($P_\mathrm{m}$) across the observing period. The solid line is a model of the Moon's antenna temperature based on the CLASS Q-band beam solid angle, the Moon's time-dependent angular diameter, the Moon's temperature oscillations that follow its phase, and an absolute calibration to a mean Moon temperature of \SI{\Tm}{\kelvin} at \SI{\centerfreqps}{\giga\hertz}~\citep{moon_calib}.
\textit{Middle}: The solid line plots the Moon's angular diameter over time. 
The 27-day elliptical Moon orbit around the Earth is perturbed by the Sun.
The dashed line plots the Moon phase delayed by \phaseoffset~days to compensate for the measured lag in Moon brightness temperature.
The 27-day orbital period and \moonperiod-day phase period interact to create the beat pattern observed in the top panel. 
\textit{Bottom}: Data points from the top panel are divided by the beam-Moon dilution factor ($\eta_\mathrm{m} = \Omega_\mathrm{m}/\Omega'$).
The solid line plots the Moon's brightness temperature model described by equation~\ref{eqn:moon}, with its parameters calibrated to match a mean temperature of \SI{\Tm}{\kelvin} marked with the dashed line.
Differences between the data points and the model may be the result of weather, instrumental systematics, and/or limitations of the Moon emission model. 
Future work will explore implementing additional data quality tools and introducing a more complex Moon thermal model. 
}
\label{fig:moon}
\end{figure*}

The electrical current signal from each TES detector is calibrated to power deposited on its bolometer island through responsivity estimates from the most recent $I$-$V$ acquisition. 
For the entire array, we find one calibration factor ($\mathrm{d}T_{\mathrm{RJ}}/\mathrm{d}P_{\gamma}$) from power deposited at the bolometer ($\mathrm{d}P_\gamma$) to antenna (Rayleigh-Jeans) temperature ($\mathrm{d}T_{\mathrm{RJ}}$) on the sky.
At $\centerfreq~\mbox{GHz}$, the conversion factor from $\mathrm{d}T_{\mathrm{RJ}}$ to CMB thermodynamic temperature ($\mathrm{d}T_{\mathrm{cmb}}$) is    $\mathrm{d}T_{\mathrm{cmb}}/\mathrm{d}T_{\mathrm{RJ}}~=~\dTcmbdTrj$.\footnote{$\mathrm{d}T_{\mathrm{cmb}}/\mathrm{d}T_{\mathrm{RJ}}\approx(e^{x_\mathrm{o}}-1)^2/x_\mathrm{o}^2 e^{x_\mathrm{o}}$ where $x_\mathrm{o} = h \nu_\mathrm{o}/k T_{\mathrm{cmb}}$}  
Individual detector TODs are calibrated to the array standard (i.e., the average) through a relative calibration factor $\epsilon_j$ equivalent to the inverse relative optical efficiency of the detector.
Hence a small $\mathrm{d}I_j$ signal of detector $j$ is calibrated to $\mathrm{d}T_{\mathrm{cmb}}$ units through
\begin{equation}
    dT_{\mathrm{cmb}} = \epsilon_j \frac{\mathrm{d}T_{\mathrm{cmb}}}{\mathrm{d}T_{\mathrm{RJ}}}\frac{\mathrm{d}T_{\mathrm{RJ}}}{\mathrm{d}P_\gamma} \frac{\mathrm{d}I_j}{S^k_j},
\end{equation}
where $k$ identifies the $I$-$V$ used to estimate the detector responsivity $S^k_j$.

The Moon is an excellent target to constrain the absolute and relative calibrations of the CLASS Q-band detectors.
At radio and millimeter wavelengths, the Moon radiates like a gray body, with frequency-dependent brightness temperature established by the optical depth of the lunar regolith and its thermal properties~\citep{linsky_moon_thermal_model,troitskii_moon_model}. 
Unlike visible light, the scattering of microwave radiation from the Sun off the Moon's surface is negligible compared to its thermal emission. 

The Moon's angular size of half a degree and $\sim\SI{200}{\kelvin}$  temperature at Q band approximates a point source for the 1.5~degree CLASS beam. 
When aligned with the beam center, the array-averaged peak Moon power measured at the bolometers is $P_\mathrm{m}\sim\SI{1.3}{\pico\watt}$, which is one-third of the average detector saturation power. 
The TES response is linear throughout the full range of the Moon signal, and a signal-to-noise ratio of $\sim\num{100000}$ is achieved.
This allows the measurement of detector pointing, beams, and calibration factors from the nominal 720$^\circ$ azimuth scan data whenever the Moon is in the field of view, increasing the observing efficiency by reducing time spent conducting targeted scans.
  
The absolute and relative detector calibrations extracted from Moon observations depend on the size of the average detector beam (see Figure \ref{fig:beam_fts_tau}) and the angular extent and brightness temperature of the Moon ($T_\mathrm{m}$) at \SI{\centerfreqps}{\giga\hertz} on the date of observation. 
Moon-centered maps indicate the average detector beam matches the full width at half maximum (FWHM) design target of 1.5 degrees (see Figure \ref{fig:beam_fts_tau}). 
The average Moon angular diameter ($D_\mathrm{m}$) of 31~arc-minutes corresponds to a beam power dilution factor ($\eta_\mathrm{m}$) given by the ratio of the moon solid angle ($\Omega_\mathrm{m}$) to the solid angle of the convolution of the beam with the moon ($\Omega'$), $\eta_\mathrm{m} =\Omega_{m}/\Omega'=$~\dilution.
This dilution factor makes the peak Moon antenna temperature $T^A_\mathrm{m} = \eta_\mathrm{m} T_\mathrm{m} \sim$\SI{16}{\kelvin}.
 
 Tidal locking of the Moon's rotation and its orbit results in one hemisphere of the Moon always facing the Earth.
 The Moon's brightness temperature averaged across its Earth-facing hemisphere and across the lunar cycle ($\overline{T}_\mathrm{m}$) has been accurately measured at Q band with the aid of an ``artificial Moon'' calibrator~\citep{Krotikov_1964,Troitsky_1968,Troitsky_1970,moon_calib}.
 At \SI{35}{\giga\hertz}, \cite{moon_calib} reports $\overline{T}_\mathrm{m} = \SI{211\pm5}{\kelvin}$, and at \SI{44}{\giga\hertz}, $\overline{T}_\mathrm{m} = \SI{208\pm5}{\kelvin}$. 
 For the CLASS \SI{\centerfreqps}{\giga\hertz} center frequency, we take $\overline{T}_\mathrm{m} = \SI{\Tm\pm5}{\kelvin}$.
 \cite{linsky_moon_center} proposed using the brightness temperature at the center of the lunar disk as a radiometric standard for wavelengths between \SI{10}{\micro\meter} and \SI{1}{\meter}.
 Near \SI{8}{\milli\meter} wavelengths, \cite{linsky_moon_center} estimates a time-averaged brightness temperature at the center of the lunar disk of $\sim\SI{230}{\kelvin}$.
 Note that the brightness temperature averaged across the entire lunar disk is lower than at the center due to colder temperatures near the poles.
 More recently, the ChangE satellite~\citep{changE} mapped the Moon temperature at \SI{37}{\giga\hertz} with high resolution; unfortunately, the absolute calibration is less reliable due to beam side-lobe pickup of the cold antenna reference \citep{changE_calib_1,changE_calib_2}.

 The temperature of the Moon's Earth-facing hemisphere oscillates with the fraction of Sun illumination or Moon phase (see bottom panel of Figure \ref{fig:moon}).
 At Q band, the maximum $T_\mathrm{m}$ peaks a few days after full Moon due to the heat capacity and thermal conductivity of the Moon's surface material~\citep{troitskii_moon_model}. 
 The Moon phase follows on average a 29.53-day cycle, while the Moon orbital period is 27.32 days.
 The Moon's elliptical orbit around Earth (strongly perturbed by the Sun) changes its angular diameter $D_\mathrm{m}$ on the sky, as shown in the middle panel of Figure \ref{fig:moon}.
 The two distinct periods of the Moon's temperature and angular diameter oscillations cause a beat pattern in the measured antenna temperature ($T^A_\mathrm{m}$, see top panel of Figure~\ref{fig:moon}). 
For example, in May 2017 the peak $T_\mathrm{m}$ coincided with minimum $D_\mathrm{m}$, nulling the fluctuation in $T^A_\mathrm{m}$, while the opposite effect occurred in November 2016 and December 2017.  

 To isolate the $T_\mathrm{m}$ oscillation from the Moon angular size variations, the $P_\mathrm{m}$ measurements across the observing era are divided by the time-dependent beam-Moon dilution factor $\eta_\mathrm{m} $: $P'_\mathrm{m} = P_\mathrm{m}/\eta_\mathrm{m}$.  
 $P'_\mathrm{m}$ data points are fit to a simple sinusoidal model over time $t$:
 \begin{equation}
     P'_\mathrm{m} =  P'_0+P'_1\cos{(2 \pi t/t_\mathrm{o} + \phi)},
     \label{eqn:moon}
 \end{equation}
where $P'_0$ is the average brightness, $P'_1$ is the amplitude of the brightness fluctuations, $t_\mathrm{o}$ the period of the oscillation, and $\phi$ the offset from full Moon. 
As expected, the fit yields $t_\mathrm{o} = 29.5$ days, the same as the Moon phase period, and $\phi = \phaseoffset$ days after full Moon, indicating a lag between full Moon illumination and maximum brightness temperature.
$P'_0$ equals \SI{\poffset}{\pico\watt}, and $P'_1$ equals \SI{\pamp}{\pico\watt} (see bottom panel of Figure~\ref{fig:moon}).
 
The Moon disk blocks the CMB radiation behind it; therefore $P'_\mathrm{m}$ measures the difference between $T_\mathrm{m}$ and the background CMB brightness temperature at \SI{\centerfreq}{\giga\hertz}, $T^B_{\mathrm{cmb}}= \SI{1.9}{\kelvin}$. 
$T^B_{\mathrm{cmb}}$ is less than the CMB's blackbody temperature $T_{\mathrm{cmb}} = \SI{2.725}{\kelvin}$~\citep{firas_2009} due to the brightness temperature definition, which is based on the Rayleigh-Jeans approximation.
The CMB temperature was not known when the ``artificial moon'' observations were made; therefore we interpret their reported average Moon temperature to be measured with respect to the CMB background :  $\overline{T}_\mathrm{m} = \langle{T}_\mathrm{m}-T^B_{\mathrm{cmb}}\rangle $.
Note that the background CMB brightness temperature is less than~1\% (and well within the uncertainty) of $\overline{T}_\mathrm{m}$.

The array's absolute calibration factor is given by the ratio of the reported average Moon brightness temperature $\overline{T}_\mathrm{m}$ and the array-averaged brightness power $P'_{0}$:
\begin{equation}
\frac{\mathrm{d}T_{\mathrm{RJ}}}{\mathrm{d}P_{\gamma}}=\frac{\overline{T}_\mathrm{m}}{P'_{0}}=\SI{\dTrjdP\pm\dTrjdPerr}{\kelvin\per\pico\watt},~\mathrm{and}
\end{equation}
\begin{equation}
\frac{\mathrm{d}T_{\mathrm{cmb}}}{\mathrm{d}P_{\gamma}}=\frac{\overline{T}_\mathrm{m}}{P'_{0}}\frac{\mathrm{d}T_{\mathrm{cmb}}}{\mathrm{d}T_{\mathrm{RJ}}}~=~\SI{\dTcmbdP\pm\dTcmbdPerr}{\kelvin\per\pico\watt}.
\end{equation}
This absolute calibration factor translates to a telescope optical efficiency of:
\begin{equation}
\eta =\left(k\Delta\nu\frac{\mathrm{d}T_{\mathrm{RJ}}}{\mathrm{d}P_{\gamma}}\right)^{-1} =  \num{\eff\pm\efferr},
\end{equation}
where the uncertainty on $\eta$ is driven by the uncertainty on $\Delta\nu$ ($\sigma_{\Delta\nu}= \SI{0.5}{\giga\hertz}$).

Relative calibration factors $\epsilon_j$ between detectors are obtained by dividing the array average Moon amplitude by the individual bolometer Moon measurement.
These relative factors account for small differences in beam solid angle, bandpass, and optical efficiency across the detector array.
Moon data indicate these are constant throughout the observing period and fall between 0.9 and 1.1. 

The $\mathrm{NEP}$ noise measurements are multiplied by $\mathrm{d}T_{\mathrm{cmb}}/\mathrm{d}P_{\gamma}$ to obtain median single detector $\mathrm{NET} = \SI{\medianDetNET}{\micro\kelvin_{\mathrm{cmb}}\sqrt{\second}}$. The average $P_{\gamma} = \SI{\popt}{\pico\watt}$ is equivalent to an antenna temperature of $T_{\gamma}~=P_{\gamma}/(\eta k \Delta \nu)~=~\SI{\Topt}{\kelvin}$. 
We estimate that \SI{5}{\kelvin} is from emission or spill within the cryostat~\citep{jeff_spie_2018} (\SI{1}{\kelvin}, \SI{4}{\kelvin}, and \SI{60}{\kelvin} filters and lenses; \SI{4}{\kelvin} cold stop; and \SI{300}{\kelvin} filters and window), \SI{6}{\kelvin} originates from the rest of the telescope (mirrors, VPM, closeout, mount enclosure, and baffle), \SI{8}{\kelvin} comes from atmospheric emission, and \SI{1.9}{\kelvin} from the CMB.
The system noise temperature, $T_{\mathrm{sys}} = \mathrm{NEP}/\eta k \sqrt{\Delta \nu} = \SI{\Tsys}{\kelvin}$,  implies an effective detector noise temperature of $T_{\mathrm{det}} = T_{\mathrm{sys}} - T_{\gamma} = \SI{\Tdet}{\kelvin}$.

\section{Tau~A intensity at Q band}
\label{sec:tauA}
\begin{table*}[ht]
\footnotesize
    \center
    \renewcommand{\arraystretch}{1.2}
    \begin{tabular}{p{1.3cm}p{0.7cm}p{0.8cm}p{1.5cm}p{1.5cm}p{1.6cm}p{5.0cm}}
    \hline
    \hline
    Instrument & Year & $\nu_e$ \newline(\SI{}{\giga\hertz}) & Flux \newline(\SI{}{\jansky})  & Flux 2017 \newline(\SI{}{\jansky})  & Flux 2017 at \newline \SI{\centerfreqtaua}{\giga\hertz}  (\SI{}{\jansky})  & References\\
    \hline
    NRL & 1966 & 31.4 & $387\pm87$ & $344\pm77$ & $321\pm73$  &
    \citet{tauA_NRL_1966} \\
    AFCRL & 1967 & 34.9 & $340^{+65}_{-40}$ & $303^{+58}_{-36}$ & $293^{+57}_{-36}$  &
    \citet{Kalaghan_1967} \\
    CBI & 2000 & 31 & $355.3\pm18$ & $341.7\pm17$ & $317\pm19$ &
    \citet{cbi_tauA_article,cbi_tauA_thesis}  \\
    VSA & 2001 & 33 & $322\pm4$ & $310\pm4$ & $294\pm11$ &
    \citet{VSA_tauA} \\
    \textit{WMAP}& 2005 & 32.96 & $342.8\pm6.4$ & $333.5\pm6.2$ & $316\pm12$ &
    \citet{wmap_weiland} \\
    \textit{WMAP}& 2005 & 40.64 & $317.7\pm8.6$ & $307.9\pm8.4$ & $314\pm13$ &
    \citet{wmap_weiland} \\
    \textit{Planck}& 2011 & 30 & $344.23\pm0.27$ & $339.51\pm0.34$ & $311\pm10$ &
    \citet{planck_tauA} \\
    \textit{Planck}& 2011 & 44 & $292.68\pm0.23$ & $288.14\pm0.41$ & $302\pm10$ &
    \citet{planck_tauA} \\
    CLASS & 2017 & \centerfreqtaua & $\tauAJy\pm\tauAJyerr$ & $\tauAJy\pm\tauAJyerr$ & $\tauAJy\pm\tauAJyerr$ &
    This paper \\
    \hline
    \hline
    \end{tabular}
    \renewcommand{\arraystretch}{1}
\caption{Table of Tau~A flux measurements between \SIlist{30;44}{\giga\hertz}.
The reported Tau~A flux measurements are
referenced to epoch 2017 by applying a per-year flux variation of \SI{-0.23\pm0.01}{\%\per yr} for measurements between \SIlist{30;36}{\giga\hertz} and \SI{-0.26\pm0.02}{\%\per yr} for measurements between \SIlist{38;44}{\giga\hertz}~\citep{wmap_weiland}. 
The uncertainty on the yearly flux variation is propagated to the 2017 flux errors.
 The measurements are converted to 2017 flux at \SI{\centerfreqtaua}{\giga\hertz} by applying the WMAP spectral model with a power index of $\num{-0.35\pm0.026}$ and propagating the model uncertainty.
In the 1960s, Tau~A was observed with the \SI{8.8}{\meter} microwave antenna at the Air Force Cambridge Laboratories (AFCRL) and with the \SI{25.9}{\meter} paraboloidal reflector at the Naval Research Laboratory's (NRL) Maryland Point Observatory.
In the early 2000s, it was observed by the Very Small Array (VSA) located at the Teide
Observatory, Izaña, Tenerife and by the Cosmic Background Imager (CBI) from the Llano de Chajnantor in northern Chile.
Measurements by the Wilkinson Microwave Anisotropy Probe (\textit{WMAP}) satellite between 2002 and 2008 are referenced to 2005, while the \textit{Planck} satellite's 2009 to 2013 observations are referenced to 2011.}
\label{table:tauA}
\end{table*}

The Crab Nebula, or Tau~A, is the remnant of supernova SN~1054.
Its spectral energy density from radio to millimeter wavelengths follows a power law emission model with spectral index $\beta ={-0.323}$~\citep{nika_tauA}. 
Flux density measurements of Tau~A between \SIlist{30;44}{\giga\hertz} are compiled in Table~\ref{table:tauA}.
Multi-year measurements from the Wilkinson Microwave Anisotropy Probe (WMAP) establish a precise model for the time and frequency dependent intensity of Tau~A between \SIlist{22;93}{\giga\hertz}~\citep{wmap_weiland}. 
This model predicts a Tau~A flux density of 312~Jy at \SI{\centerfreqtaua}{\giga\hertz} referenced to epoch 2017.

We extract a 6$\times$6~square-degree intensity map centered at Tau~A from preliminary per-detector constant elevation scan (CES) maps covering $\fsky\%$ of the sky. 
These maps contain 72 CES that are 10 to 23 hours long.
We generate simulated maps based on the \textit{WMAP} Q-band intensity map, that incorporate the CLASS beam, scan strategy, and TOD filtering.
The simulations indicate that the peak Tau~A amplitude is reduced by 5-6\% in the preliminary CLASS maps due to the high-pass filter applied to the TODs.
This bias is corrected, and the results from the 41 detectors that point low enough on the sky to observe Tau~A are averaged.

Tau A's  7$\times$5~arcmin$^2$   \citep{green_tauA} or $\SI{3}{\micro\steradian}$ angular extent makes it effectively a point source when compared to the Q-band beam solid angle ($\Omega = \SI{\solidangle\pm\sderr}{\micro\steradian}$).
The CLASS map of Tau A is consistent with the CLASS beam derived from moon measurements ($\mathrm{FWHM}\approx\SI{1.5}{\degree}$, see Figure~\ref{fig:beam_fts_tau}) with a peak amplitude of $T_{\mathrm{A}}=\SI{\tauAT\pm\tauATerr}{\milli\kelvin}$.
 For the CLASS Q-band instrument, the peak amplitude in antenna temperature (K) is converted to spectral flux density (Jy) for an unresolved (i.e., point) source through the factor \citep{page_2003,jarosik_2011}
\begin{equation}
    \Gamma = \frac{c^2}{2 k \Omega \nu^2_{e}} = \SI{\pointfluxtauA\pm\pointfluxerr}{\micro\kelvin\per\jansky},
\end{equation}
where the effective central frequency of Tau~A across the CLASS Q bandpass is $\nu_{e} = \SI{\centerfreqtaua\pm\centerfreqerr}{\giga\hertz}$.\footnote{Effective frequency for a point source is defined as $\nu_e = \int \nu \nu^{\alpha} f(\nu) d\nu/ \int \nu^{\alpha} f(\nu) d\nu$, where $f(\nu)$ describes the instrument response (passband etc), $\alpha$ parametrizes the point source flux density $S=S_e(\nu/\nu_e)^{-\alpha}$, and $\alpha=-0.3$ for Tau A \citep{page2003}.} 

\begin{figure}
\center
\includegraphics[width = 0.47\textwidth]{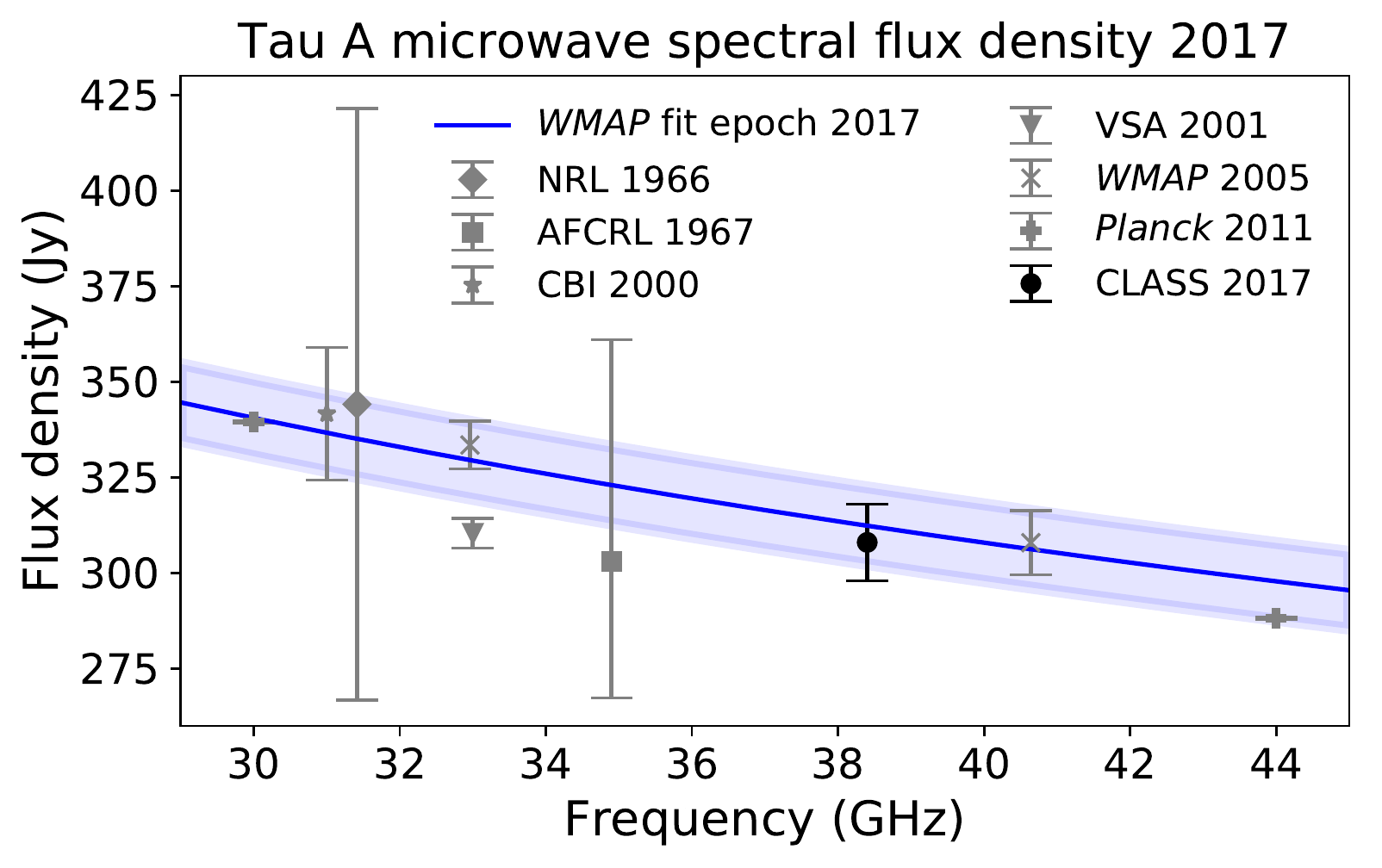}
\caption{Tau~A flux density measurements found in Table \ref{table:tauA} and referenced to epoch 2017.
The solid line is the \textit{WMAP} \SI{22}{\giga\hertz} to \SI{93}{\giga\hertz} Tau~A flux density vs frequency model: log~S(Jy)~=~$2.502-0.350$~log($\nu$/\SI{40}{\giga\hertz}) \citep{wmap_weiland}, referenced to epoch 2017.
The shaded region is the 1$\sigma$ contour of the model's flux prediction including spectral and time evolution uncertainty.
The CLASS intensity map calibrated through the Moon yields a Tau~A flux density of $\SI{\tauAJy\pm\tauAJyerr}{Jy}$ at $\nu_{e} = \SI{\centerfreqtaua\pm\centerfreqerr}{\giga\hertz}$.
The \SI{\solidangle}{\micro\steradian} CLASS beam solid angle dilutes Tau~A to an antenna temperature of \SI{\tauAT\pm\tauATerr}{\milli\kelvin}.
}
\label{fig:tauA}
\end{figure}

Dividing the peak temperature measured for Tau~A by $\Gamma$ gives a flux density of $\tauAJy\pm\tauAJyerr$~Jy. 
Figure \ref{fig:tauA} plots the flux density measurements tabulated in Table~\ref{table:tauA}.
The \textit{WMAP} Tau~A flux density model, which includes a yearly rate of decline and a spectral index, matches well with the measurements in the \SIlist{30;44}{\giga\hertz} frequency range.
Note that the reported CLASS Tau~A flux density is independent of CMB calibration and rather is anchored to the Moon brightness temperature. 
In other words, the Tau~A measurement shows that the CLASS antenna temperature calibration based on Moon observations is consistent with the \textit{WMAP} calibration based on the CMB dipole. 
\section{Conclusion}
\label{sec:discussion}

In this paper, we have established the basic on-sky performance of the CLASS telescopes.
The stability and time constants of the Q-band TES bolometers are within specification.
The array average \SI{\popt}{\pico\watt} optical loading, \SI{\nep}{\atto\watt\sqrt{\second}} NEP, and \SI{\Tsys}{\kelvin} system noise temperature satisfy the design targets.
A \SI{\dTrjdP\pm\dTrjdPerr}{\kelvin\per\pico\watt} calibration factor that converts from optical power measured at the bolometer to Rayleigh-Jeans temperature on the sky is obtained from fitting hundreds of Moon observations to a Moon brightness temperature model that follows the Moon's orbit and phase.
This calibration factor translates to a telescope optical efficiency of $\eff\pm\efferr$ and is used to construct a Tau~A intensity map from the nominal CMB scans.
We report a Tau~A flux density of \SI{\tauAJy\pm\tauAJyerr}{Jy} at \SI{\centerfreqtaua\pm\centerfreqerr}{\giga\hertz}, consistent with the \textit{WMAP} Tau~A time-dependent spectral flux density model.
The 1$\sigma$ error of the CLASS measurement includes the uncertainty in the bandpass center frequency, the calibration to antenna temperature, and the Tau~A peak amplitude.

Between June 2016 and March 2018, CLASS carried out the largest ground-based Q-band CMB sky survey to date, covering $\fsky\%$ of the sky. 
Comparable large-scale ground-based surveys at low-frequencies include \cite{quijote_2017} and \cite{cbass_2018}.
During this initial CLASS observing campaign, \ndet~Q-band bolometers were optically sensitive, with a median per detector NET of \SI{\medianDetNET}{\micro\kelvin_{\mathrm{cmb}}\sqrt{\second}}, which implies a median instantaneous array sensitivity of \SI{\arraynet}{\micro\kelvin_{\mathrm{cmb}}\sqrt{\second}}.
For comparison, the combined polarization sensitivity of the four \textit{WMAP} \SI{41}{\giga\hertz} radiometers was \SI{469}{\micro\kelvin_{\mathrm{cmb}}\sqrt{\second}}~\citep{jarosik_2003}, and the combined sensitivity of the six \textit{Planck} \SI{44}{\giga\hertz} radiometers was \SI{174}{\micro\kelvin_{\mathrm{cmb}}\sqrt{\second}}~\citep{planck_2015_results_II}.

This is the first of a series of papers to be published on the initial two years of CLASS \SI{40}{\giga\hertz} observations. 
Additional articles will present the CLASS beam and window function, CLASS polarization modulation and stability, CMB survey maps, and science results derived therefrom. 
Beyond the first two years, the \SI{40}{\giga\hertz}  telescope continues to acquire data together with a \SI{90}{\giga\hertz} telescope that was installed in 2018.
An additional telescope at \SI{90}{\giga\hertz} and a 150/220 dichroic telescope will be installed in the near future.
The nominal survey ends in late 2021 with plans for extensions thereon.

\section*{Acknowledgements}

We acknowledge the National Science Foundation Division of Astronomical Sciences for their support of CLASS under Grant Numbers 0959349, 1429236, 1636634, and 1654494. The CLASS project employs detector technology developed under several previous and ongoing NASA grants. Detector development work at JHU was funded by NASA grant number NNX14AB76A. K. Harrington is supported by NASA Space Technology Research Fellowship grant number NX14AM49H. Basti\'an Pradenas is supported by the Fondecyt Regular Project No. 1171811 (CONICYT) and CONICYT-PFCHA Magister Nacional Scholarship 2016-22161360. We thank the anonymous reviewer for his or her careful reading of our manuscript and many insightful comments and suggestions. We thank Philip Mauskopf for useful discussions of bolometer noise. We acknowledge important contributions from Keisuke Osumi, Mark Halpern, Mandana Amiri, Gary Rhodes, Janet Weiland, Keisuke Osumi, Mario Aguilar, Yunyang Li, Isu Ravi, Tiffany Wei, Connor Henley, Max Abitbol, Lindsay Lowry, and Fletcher Boone. We thank William Deysher, Maria Jose Amaral, and Chantal Boisvert for administrative support. We acknowledge productive collaboration with Dean Carpenter and the JHU Physical Sciences Machine Shop team. Part of this research project was conducted using computational resources at the Maryland Advanced Research Computing Center (MARCC). Some of
the results in this paper have been derived using the HEALPix package \citep{healpix}. We further acknowledge the very generous support of Jim and Heather Murren (JHU A\&S '88), Matthew Polk (JHU A\&S Physics BS '71), David Nicholson, and Michael Bloomberg (JHU Engineering '64). CLASS is located in the Parque Astron\'omico Atacama in northern Chile under the auspices of the Comisi\'on Nacional de Investigaci\'on Científica y Tecnológica de Chile (CONICYT). R.D. and P.F. thank CONICYT for grants Anillo ACT-1417, QUIMAL 160009, and BASAL AFB-170002.


\bibliography{Qband_onsky}

\begin{thebibliography}{}
\expandafter\ifx\csname natexlab\endcsname\relax\def\natexlab#1{#1}\fi

\bibitem[{Ade {et~al.}(2009)Ade, Chuss, Hanany, Haynes, Keating, Kogut, Ruhl,
  Pisano, Savini, \& Wollack}]{wollack_omt}
Ade, P. A.~R., Chuss, D.~T., Hanany, S., {et~al.} 2009, Journal of Physics:
  Conference Series, 155, 012006

\bibitem[{{Albrecht} \& {Steinhardt}(1982)}]{albrecht/steinhardt:1982}
{Albrecht}, A., \& {Steinhardt}, P.~J. 1982, Physical Review Letters, 48, 1220

\bibitem[{{Allison} {et~al.}(2015){Allison}, {Caucal}, {Calabrese}, {Dunkley},
  \& {Louis}}]{Allison2015}
{Allison}, R., {Caucal}, P., {Calabrese}, E., {Dunkley}, J., \& {Louis}, T.
  2015, \prd, 92, 123535

\bibitem[{{Appel} {et~al.}(2014){Appel}, {Ali}, {Amiri}, {Araujo}, {Bennet},
  {Boone}, {Chan}, {Cho}, {Chuss}, {Colazo}, {Crowe}, {Denis}, {D{\"u}nner},
  {Eimer}, {Essinger-Hileman}, {Gothe}, {Halpern}, {Harrington}, {Hilton},
  {Hinshaw}, {Huang}, {Irwin}, {Jones}, {Karakula}, {Kogut}, {Larson}, {Limon},
  {Lowry}, {Marriage}, {Mehrle}, {Miller}, {Miller}, {Moseley}, {Novak},
  {Reintsema}, {Rostem}, {Stevenson}, {Towner}, {U-Yen}, {Wagner}, {Watts},
  {Wollack}, {Xu}, \& {Zeng}}]{spie_jappel}
{Appel}, J.~W., {Ali}, A., {Amiri}, M., {et~al.} 2014, in \procspie, Vol. 9153,
  Millimeter, Submillimeter, and Far-Infrared Detectors and Instrumentation for
  Astronomy VII, 91531J

\bibitem[{{Battistelli} {et~al.}(2008){Battistelli}, {Amiri}, {Burger},
  {Halpern}, {Knotek}, {Ellis}, {Gao}, {Kelly}, {Macintosh}, {Irwin}, \&
  {Reintsema}}]{ubc_mce}
{Battistelli}, E.~S., {Amiri}, M., {Burger}, B., {et~al.} 2008, Journal of Low
  Temperature Physics, 151, 908

\bibitem[{{Bennett} {et~al.}(2013){Bennett}, {Larson}, {Weiland}, {Jarosik},
  {Hinshaw}, {Odegard}, {Smith}, {Hill}, {Gold}, {Halpern}, {Komatsu}, {Nolta},
  {Page}, {Spergel}, {Wollack}, {Dunkley}, {Kogut}, {Limon}, {Meyer}, {Tucker},
  \& {Wright}}]{bennett:2013}
{Bennett}, C.~L., {Larson}, D., {Weiland}, J.~L., {et~al.} 2013, \apjs, 208, 20

\bibitem[{{BICEP2 Collaboration} {et~al.}(2016){BICEP2 Collaboration}, {Keck
  Array Collaboration}, {Ade}, {Ahmed}, {Aikin}, {Alexander}, {Barkats},
  {Benton}, {Bischoff}, {Bock}, {Bowens-Rubin}, {Brevik}, {Buder}, {Bullock},
  {Buza}, {Connors}, {Crill}, {Duband}, {Dvorkin}, {Filippini}, {Fliescher},
  {Grayson}, {Halpern}, {Harrison}, {Hilton}, {Hui}, {Irwin}, {Karkare},
  {Karpel}, {Kaufman}, {Keating}, {Kefeli}, {Kernasovskiy}, {Kovac}, {Kuo},
  {Leitch}, {Lueker}, {Megerian}, {Netterfield}, {Nguyen}, {O'Brient},
  {Ogburn}, {Orlando}, {Pryke}, {Richter}, {Schwarz}, {Sheehy}, {Staniszewski},
  {Steinbach}, {Sudiwala}, {Teply}, {Thompson}, {Tolan}, {Tucker}, {Turner},
  {Vieregg}, {Weber}, {Wiebe}, {Willmert}, {Wong}, {Wu}, \& {Yoon}}]{bicep2016}
{BICEP2 Collaboration}, {Keck Array Collaboration}, {Ade}, P.~A.~R., {et~al.}
  2016, Physical Review Letters, 116, 031302

\bibitem[{{BICEP2 Collaboration} {et~al.}(2018){BICEP2 Collaboration}, {Keck
  Array Collaboration}, {Ade}, {Ahmed}, {Aikin}, {Alexander}, {Barkats},
  {Benton}, {Bischoff}, {Bock}, {Bowens-Rubin}, {Brevik}, {Buder}, {Bullock},
  {Buza}, {Connors}, {Cornelison}, {Crill}, {Crumrine}, {Dierickx}, {Duband},
  {Dvorkin}, {Filippini}, {Fliescher}, {Grayson}, {Hall}, {Halpern},
  {Harrison}, {Hildebrandt}, {Hilton}, {Hui}, {Irwin}, {Kang}, {Karkare},
  {Karpel}, {Kaufman}, {Keating}, {Kefeli}, {Kernasovskiy}, {Kovac}, {Kuo},
  {Larsen}, {Lau}, {Leitch}, {Lueker}, {Megerian}, {Moncelsi}, {Namikawa},
  {Netterfield}, {Nguyen}, {O'Brient}, {Ogburn}, {Palladino}, {Pryke},
  {Racine}, {Richter}, {Schillaci}, {Schwarz}, {Sheehy}, {Soliman},
  {St.~Germaine}, {Staniszewski}, {Steinbach}, {Sudiwala}, {Teply}, {Thompson},
  {Tolan}, {Tucker}, {Turner}, {Umilt{\`a}}, {Vieregg}, {Wandui}, {Weber},
  {Wiebe}, {Willmert}, {Wong}, {Wu}, {Yang}, {Yoon}, \&
  {Zhang}}]{bicep_keck_2018}
---. 2018, Physical Review Letters, 121, 221301

\bibitem[{{Bustos} {et~al.}(2014){Bustos}, {Rubio}, {Ot{\'a}rola}, \&
  {Nagar}}]{parque_atacama}
{Bustos}, R., {Rubio}, M., {Ot{\'a}rola}, A., \& {Nagar}, N. 2014, Publications
  of the Astronomical Society of the Pacific, 126, 1126

\bibitem[{{Cartwright}(2003)}]{cbi_tauA_thesis}
{Cartwright}, J.~K. 2003, PhD thesis, California Institute of Technology

\bibitem[{{Cartwright} {et~al.}(2005){Cartwright}, {Pearson}, {Readhead},
  {Shepherd}, {Sievers}, \& {Taylor}}]{cbi_tauA_article}
{Cartwright}, J.~K., {Pearson}, T.~J., {Readhead}, A.~C.~S., {et~al.} 2005,
  \apj, 623, 11

\bibitem[{{Chuss} {et~al.}(2012{\natexlab{a}}){Chuss}, {Wollack}, {Henry},
  {Hui}, {Juarez}, {Krejny}, {Moseley}, \& {Novak}}]{chuss_vpm_2012}
{Chuss}, D.~T., {Wollack}, E.~J., {Henry}, R., {et~al.} 2012{\natexlab{a}},
  \ao, 51, 197

\bibitem[{{Chuss} {et~al.}(2012{\natexlab{b}}){Chuss}, {Bennett}, {Costen},
  {Crowe}, {Denis}, {Eimer}, {Lourie}, {Marriage}, {Moseley}, {Rostem},
  {Stevenson}, {Towner}, {U-Yen}, {Voellmer}, {Wollack}, \&
  {Zeng}}]{dchuss_qdet_2012}
{Chuss}, D.~T., {Bennett}, C.~L., {Costen}, N., {et~al.} 2012{\natexlab{b}},
  Journal of Low Temperature Physics, 167, 923

\bibitem[{{Chuss} {et~al.}(2014){Chuss}, {Ali}, {Appel}, {Bennett}, {Colazo},
  {Crowe}, {Denis}, {Eimer}, {Essinger-Hileman}, {Marriage}, {Moseley},
  {Rostem}, {Stevenson}, {Towner}, {U-Yen}, {Wollack}, \&
  {Zeng}}]{dchuss_qdet_2014}
{Chuss}, D.~T., {Ali}, A., {Appel}, J.~W., {et~al.} 2014, in American
  Astronomical Society Meeting Abstracts \#223, Vol. 223, 439.03

\bibitem[{{Dahal} {et~al.}(2018){Dahal}, {Ali}, {Appel}, {Essinger-Hileman},
  {Bennett}, {Brewer}, {Bustos}, {Chan}, {Chuss}, {Cleary}, {Colazo}, {Couto},
  {Denis}, {D{\"u}nner}, {Eimer}, {Engelhoven}, {Fluxa}, {Halpern},
  {Harrington}, {Helson}, {Hilton}, {Hinshaw}, {Hubmayr}, {Iuliano}, {Karakla},
  {Marriage}, {McMahon}, {Miller}, {Nu{\~n}ez}, {Padilla}, {Palma}, {Parker},
  {Petroff}, {Pradenas}, {Reeves}, {Reintsema}, {Rostem}, {Sagliocca}, {U-Yen},
  {Valle}, {Wang}, {Wang}, {Watts}, {Weiland}, {Wollack}, {Xu}, {Yan}, \&
  {Zeng}}]{sumit_spie_2018}
{Dahal}, S., {Ali}, A., {Appel}, J.~W., {et~al.} 2018, in Society of
  Photo-Optical Instrumentation Engineers (SPIE) Conference Series, Vol. 10708,
  Society of Photo-Optical Instrumentation Engineers (SPIE) Conference Series,
  107081Y

\bibitem[{{Denis} {et~al.}(2009){Denis}, {Cao}, {Chuss}, {Eimer}, {Hinderks},
  {Hsieh}, {Moseley}, {Stevenson}, {Talley}, {U. -yen}, \&
  {Wollack}}]{denis_fabq}
{Denis}, K.~L., {Cao}, N.~T., {Chuss}, D.~T., {et~al.} 2009, in American
  Institute of Physics Conference Series, ed. B.~{Young}, B.~{Cabrera}, \&
  A.~{Miller}, Vol. 1185, 371--374

\bibitem[{{Denis} {et~al.}(2016){Denis}, {Ali}, {Appel}, {Bennett}, {Chang},
  {Chuss}, {Colazo}, {Costen}, {Essinger-Hileman}, {Hu}, {Marriage}, {Rostem},
  {U-Yen}, \& {Wollack}}]{denis_fabw}
{Denis}, K.~L., {Ali}, A., {Appel}, J., {et~al.} 2016, Journal of Low
  Temperature Physics, 184, 668

\bibitem[{{Doriese} {et~al.}(2016){Doriese}, {Morgan}, {Bennett}, {Denison},
  {Fitzgerald}, {Fowler}, {Gard}, {Hays-Wehle}, {Hilton}, {Irwin}, {Joe},
  {Mates}, {O'Neil}, {Reintsema}, {Robbins}, {Schmidt}, {Swetz}, {Tatsuno},
  {Vale}, \& {Ullom}}]{nist_tdm_mux13b}
{Doriese}, W.~B., {Morgan}, K.~M., {Bennett}, D.~A., {et~al.} 2016, Journal of
  Low Temperature Physics, 184, 389

\bibitem[{{Eimer} {et~al.}(2012){Eimer}, {Bennett}, {Chuss}, {Marriage},
  {Wollack}, \& {Zeng}}]{joseph_SPIE}
{Eimer}, J.~R., {Bennett}, C.~L., {Chuss}, D.~T., {et~al.} 2012, in Society of
  Photo-Optical Instrumentation Engineers (SPIE) Conference Series, Vol. 8452,
  Society of Photo-Optical Instrumentation Engineers (SPIE) Conference Series

\bibitem[{{Essinger-Hileman} {et~al.}(2014){Essinger-Hileman}, {Ali}, {Amiri},
  {Appel}, {Araujo}, {Bennett}, {Boone}, {Chan}, {Cho}, {Chuss}, {Colazo},
  {Crowe}, {Denis}, {D{\"u}nner}, {Eimer}, {Gothe}, {Halpern}, {Harrington},
  {Hilton}, {Hinshaw}, {Huang}, {Irwin}, {Jones}, {Karakla}, {Kogut}, {Larson},
  {Limon}, {Lowry}, {Marriage}, {Mehrle}, {Miller}, {Miller}, {Moseley},
  {Novak}, {Reintsema}, {Rostem}, {Stevenson}, {Towner}, {U-Yen}, {Wagner},
  {Wollack}, {Xu}, \& {Zeng}}]{tom_spie}
{Essinger-Hileman}, T., {Ali}, A., {Amiri}, M., {et~al.} 2014, in SPIE, Vol.
  915354, Millimeter, Submillimeter, and Far-Infrared Detectors and
  Instrumentation for Astronomy VII

\bibitem[{Fixsen(2009)}]{firas_2009}
Fixsen, D.~J. 2009, \apj, 707, 916

\bibitem[{{G{\'e}nova-Santos} {et~al.}(2017){G{\'e}nova-Santos},
  {Rubi{\~n}o-Mart{\'{\i}}n}, {Pel{\'a}ez-Santos}, {Poidevin}, {Rebolo},
  {Vignaga}, {Artal}, {Harper}, {Hoyland}, {Lasenby},
  {Mart{\'{\i}}nez-Gonz{\'a}lez}, {Piccirillo}, {Tramonte}, \&
  {Watson}}]{quijote_2017}
{G{\'e}nova-Santos}, R., {Rubi{\~n}o-Mart{\'{\i}}n}, J.~A.,
  {Pel{\'a}ez-Santos}, A., {et~al.} 2017, \mnras, 464, 4107

\bibitem[{{G{\'o}rski} {et~al.}(2005){G{\'o}rski}, {Hivon}, {Banday},
  {Wandelt}, {Hansen}, {Reinecke}, \& {Bartelmann}}]{healpix}
{G{\'o}rski}, K.~M., {Hivon}, E., {Banday}, A.~J., {et~al.} 2005, \apj, 622,
  759

\bibitem[{{Green}(2009)}]{green_tauA}
{Green}, D.~A. 2009, Bulletin of the Astronomical Society of India, 37, 45

\bibitem[{{Guth}(1981)}]{guth:1981}
{Guth}, A.~H. 1981, \prd, 23, 347

\bibitem[{{Hafez} {et~al.}(2008){Hafez}, {Davies}, {Davis}, {Dickinson},
  {Battistelli}, {Blanco}, {Cleary}, {Franzen}, {Genova-Santos}, {Grainge},
  {Hobson}, {Jones}, {Lancaster}, {Lasenby}, {Padilla-Torres},
  {Rubi{\~n}o-Martin}, {Rebolo}, {Saunders}, {Scott}, {Taylor}, {Titterington},
  {Tucci}, \& {Watson}}]{VSA_tauA}
{Hafez}, Y.~A., {Davies}, R.~D., {Davis}, R.~J., {et~al.} 2008, \mnras, 388,
  1775

\bibitem[{{Harrington} {et~al.}(2016){Harrington}, {Marriage}, {Ali}, {Appel},
  {Bennett}, {Boone}, {Brewer}, {Chan}, {Chuss}, {Colazo}, {Dahal}, {Denis},
  {D{\"u}nner}, {Eimer}, {Essinger-Hileman}, {Fluxa}, {Halpern}, {Hilton},
  {Hinshaw}, {Hubmayr}, {Iuliano}, {Karakla}, {McMahon}, {Miller}, {Moseley},
  {Palma}, {Parker}, {Petroff}, {Pradenas}, {Rostem}, {Sagliocca}, {Valle},
  {Watts}, {Wollack}, {Xu}, \& {Zeng}}]{harrington2016}
{Harrington}, K., {Marriage}, T., {Ali}, A., {et~al.} 2016, in \procspie, Vol.
  9914, Millimeter, Submillimeter, and Far-Infrared Detectors and
  Instrumentation for Astronomy VIII, 99141K

\bibitem[{{Harrington} {et~al.}(2018){Harrington}, {Eimer}, {Chuss}, {Petroff},
  {Cleary}, {DeGeorge}, {Grunberg}, {Ali}, {Appel}, {Bennett}, {Brewer},
  {Bustos}, {Chan}, {Couto}, {Dahal}, {Denis}, {D{\"u}nner},
  {Essinger-Hileman}, {Fluxa}, {Halpern}, {Hilton}, {Hinshaw}, {Hubmayr},
  {Iuliano}, {Karakla}, {Marriage}, {McMahon}, {Miller}, {Nu{\~n}ez},
  {Padilla}, {Palma}, {Parker}, {Pradenas Marquez}, {Reeves}, {Reintsema},
  {Rostem}, {Augusto Nunes Valle}, {Van Engelhoven}, {Wang}, {Wang}, {Watts},
  {Weiland}, {Wollack}, {Xu}, {Yan}, \& {Zeng}}]{Harrington2018}
{Harrington}, K., {Eimer}, J., {Chuss}, D.~T., {et~al.} 2018, in Society of
  Photo-Optical Instrumentation Engineers (SPIE) Conference Series, Vol. 10708,
  Society of Photo-Optical Instrumentation Engineers (SPIE) Conference Series,
  107082M

\bibitem[{{Henning} {et~al.}(2018){Henning}, {Sayre}, {Reichardt}, {Ade},
  {Anderson}, {Austermann}, {Beall}, {Bender}, {Benson}, {Bleem}, {Carlstrom},
  {Chang}, {Chiang}, {Cho}, {Citron}, {Corbett Moran}, {Crawford}, {Crites},
  {de Haan}, {Dobbs}, {Everett}, {Gallicchio}, {George}, {Gilbert},
  {Halverson}, {Harrington}, {Hilton}, {Holder}, {Holzapfel}, {Hoover}, {Hou},
  {Hrubes}, {Huang}, {Hubmayr}, {Irwin}, {Keisler}, {Knox}, {Lee}, {Leitch},
  {Li}, {Lowitz}, {Manzotti}, {McMahon}, {Meyer}, {Mocanu}, {Montgomery},
  {Nadolski}, {Natoli}, {Nibarger}, {Novosad}, {Padin}, {Pryke}, {Ruhl},
  {Saliwanchik}, {Schaffer}, {Sievers}, {Smecher}, {Stark}, {Story}, {Tucker},
  {Vanderlinde}, {Veach}, {Vieira}, {Wang}, {Whitehorn}, {Wu}, \&
  {Yefremenko}}]{spt_2018}
{Henning}, J.~W., {Sayre}, J.~T., {Reichardt}, C.~L., {et~al.} 2018, \apj, 852,
  97

\bibitem[{{Hinshaw} {et~al.}(2013){Hinshaw}, {Larson}, {Komatsu}, {Spergel},
  {Bennett}, {Dunkley}, {Nolta}, {Halpern}, {Hill}, {Odegard}, {Page}, {Smith},
  {Weiland}, {Gold}, {Jarosik}, {Kogut}, {Limon}, {Meyer}, {Tucker}, {Wollack},
  \& {Wright}}]{Hinshaw2013}
{Hinshaw}, G., {Larson}, D., {Komatsu}, E., {et~al.} 2013, \apjs, 208, 19

\bibitem[{{Hobbs} {et~al.}(1968){Hobbs}, {Corbett}, \&
  {Santini}}]{tauA_NRL_1966}
{Hobbs}, R.~W., {Corbett}, H.~H., \& {Santini}, N.~J. 1968, \apj, 152, 43

\bibitem[{{Hu} {et~al.}(2017){Hu}, {Chan}, {Zheng}, {Tsang}, \&
  {Xu}}]{changE_calib_1}
{Hu}, G.-P., {Chan}, K.~L., {Zheng}, Y.-C., {Tsang}, K.~T., \& {Xu}, A.-A.
  2017, \icarus, 294, 72

\bibitem[{Irwin \& Hilton(2005)}]{irwin_hilton}
Irwin, K., \& Hilton, G. 2005, in Topics in Applied Physics, Vol.~99, Cryogenic
  Particle Detection, ed. C.~Enss (Springer Berlin / Heidelberg), 81--97

\bibitem[{{Iuliano} {et~al.}(2018){Iuliano}, {Eimer}, {Parker}, {Rhoades},
  {Ali}, {Appel}, {Bennett}, {Brewer}, {Bustos}, {Chuss}, {Cleary}, {Couto},
  {Dahal}, {Denis}, {D{\"u}nner}, {Essinger-Hileman}, {Fluxa}, {Halpern},
  {Harrington}, {Helson}, {Hilton}, {Hinshaw}, {Hubmayr}, {Karakla},
  {Marriage}, {Miller}, {McMahon}, {Nu{\~n}ez}, {Padilla}, {Palma}, {Petroff},
  {Pradenas M{\'a}rquez}, {Reeves}, {Reintsema}, {Rostem}, {Augusto Nunes
  Valle}, {Van Engelhoven}, {Wang}, {Wang}, {Watts}, {Weiland}, {Wollack},
  {Xu}, {Yan}, \& {Zeng}}]{jeff_spie_2018}
{Iuliano}, J., {Eimer}, J., {Parker}, L., {et~al.} 2018, in Society of
  Photo-Optical Instrumentation Engineers (SPIE) Conference Series, Vol. 10708,
  Society of Photo-Optical Instrumentation Engineers (SPIE) Conference Series,
  1070828

\bibitem[{{Jarosik} {et~al.}(2003){Jarosik}, {Barnes}, {Bennett}, {Halpern},
  {Hinshaw}, {Kogut}, {Limon}, {Meyer}, {Page}, {Spergel}, {Tucker}, {Weiland},
  {Wollack}, \& {Wright}}]{jarosik_2003}
{Jarosik}, N., {Barnes}, C., {Bennett}, C.~L., {et~al.} 2003, \apjs, 148, 29

\bibitem[{{Jarosik} {et~al.}(2011){Jarosik}, {Bennett}, {Dunkley}, {Gold},
  {Greason}, {Halpern}, {Hill}, {Hinshaw}, {Kogut}, {Komatsu}, {Larson},
  {Limon}, {Meyer}, {Nolta}, {Odegard}, {Page}, {Smith}, {Spergel}, {Tucker},
  {Weiland}, {Wollack}, \& {Wright}}]{jarosik_2011}
{Jarosik}, N., {Bennett}, C.~L., {Dunkley}, J., {et~al.} 2011, \apjs, 192, 14

\bibitem[{{Jones} {et~al.}(2018){Jones}, {Taylor}, {Aich}, {Copley}, {Chiang},
  {Davis}, {Dickinson}, {Grumitt}, {Hafez}, {Heilgendorff}, {Holler}, {Irfan},
  {Jew}, {John}, {Jonas}, {King}, {Leahy}, {Leech}, {Leitch}, {Muchovej},
  {Pearson}, {Peel}, {Readhead}, {Sievers}, {Stevenson}, \&
  {Zuntz}}]{cbass_2018}
{Jones}, M.~E., {Taylor}, A.~C., {Aich}, M., {et~al.} 2018, \mnras, 480, 3224

\bibitem[{{Kalaghan} \& {Wulfsberg}(1967)}]{Kalaghan_1967}
{Kalaghan}, P.~M., \& {Wulfsberg}, K.~N. 1967, \aj, 72, 1051

\bibitem[{{Kamionkowski} {et~al.}(1997){Kamionkowski}, {Kosowsky}, \&
  {Stebbins}}]{kamionkowski}
{Kamionkowski}, M., {Kosowsky}, A., \& {Stebbins}, A. 1997, \prd, 55, 7368

\bibitem[{{Kovac} {et~al.}(2002){Kovac}, {Leitch}, {Pryke}, {Carlstrom},
  {Halverson}, \& {Holzapfel}}]{dasi}
{Kovac}, J.~M., {Leitch}, E.~M., {Pryke}, C., {et~al.} 2002, \nat, 420, 772

\bibitem[{{Krotikov} \& {Pelyushenko}(1987)}]{moon_calib}
{Krotikov}, V.~D., \& {Pelyushenko}, S.~A. 1987, \sovast, 31, 216

\bibitem[{{Krotikov} \& {Troitski{\u i}}(1964)}]{Krotikov_1964}
{Krotikov}, V.~D., \& {Troitski{\u i}}, V.~S. 1964, Soviet Physics Uspekhi, 6,
  841

\bibitem[{{Kusaka} {et~al.}(2014){Kusaka}, {Essinger-Hileman}, {Appel},
  {Gallardo}, {Irwin}, {Jarosik}, {Nolta}, {Page}, {Parker}, {Raghunathan},
  {Sievers}, {Simon}, {Staggs}, \& {Visnjic}}]{abs_hwp}
{Kusaka}, A., {Essinger-Hileman}, T., {Appel}, J.~W., {et~al.} 2014, Review of
  Scientific Instruments, 85, 024501

\bibitem[{{Kusaka} {et~al.}(2018){Kusaka}, {Appel}, {Essinger-Hileman},
  {Beall}, {Campusano}, {Cho}, {Choi}, {Crowley}, {Fowler}, {Gallardo},
  {Hasselfield}, {Hilton}, {Ho}, {Irwin}, {Jarosik}, {Niemack}, {Nixon},
  {\~{}Nolta}, {Page}, {Palma}, {Parker}, {Raghunathan}, {Reintsema},
  {Sievers}, {Simon}, {Staggs}, {Visnjic}, \& {Yoon}}]{abs_final}
{Kusaka}, A., {Appel}, J., {Essinger-Hileman}, T., {et~al.} 2018, \jcap, 9, 005

\bibitem[{{Linde}(1982)}]{linde:1982}
{Linde}, A.~D. 1982, Physics Letters B, 108, 389

\bibitem[{{Linsky}(1966)}]{linsky_moon_thermal_model}
{Linsky}, J.~L. 1966, \icarus, 5, 606

\bibitem[{{Linsky}(1973)}]{linsky_moon_center}
---. 1973, \apjs, 25, 163

\bibitem[{{Louis} {et~al.}(2017){Louis}, {Grace}, {Hasselfield}, {Lungu},
  {Maurin}, {Addison}, {Ade}, {Aiola}, {Allison}, {Amiri}, {Angile},
  {Battaglia}, {Beall}, {de Bernardis}, {Bond}, {Britton}, {Calabrese}, {Cho},
  {Choi}, {Coughlin}, {Crichton}, {Crowley}, {Datta}, {Devlin}, {Dicker},
  {Dunkley}, {D{\"u}nner}, {Ferraro}, {Fox}, {Gallardo}, {Gralla}, {Halpern},
  {Henderson}, {Hill}, {Hilton}, {Hilton}, {Hincks}, {Hlozek}, {Ho}, {Huang},
  {Hubmayr}, {Huffenberger}, {Hughes}, {Infante}, {Irwin}, {Muya Kasanda},
  {Klein}, {Koopman}, {Kosowsky}, {Li}, {Madhavacheril}, {Marriage}, {McMahon},
  {Menanteau}, {Moodley}, {Munson}, {Naess}, {Nati}, {Newburgh}, {Nibarger},
  {Niemack}, {Nolta}, {Nu{\~n}ez}, {Page}, {Pappas}, {Partridge}, {Rojas},
  {Schaan}, {Schmitt}, {Sehgal}, {Sherwin}, {Sievers}, {Simon}, {Spergel},
  {Staggs}, {Switzer}, {Thornton}, {Trac}, {Treu}, {Tucker}, {Van Engelen},
  {Ward}, \& {Wollack}}]{actpol_2017}
{Louis}, T., {Grace}, E., {Hasselfield}, M., {et~al.} 2017, \jcap, 6, 031

\bibitem[{{Martin} \& {Puplett}(1970)}]{mp_fts}
{Martin}, D.~H., \& {Puplett}, E. 1970, Infrared Physics, 10, 105

\bibitem[{Mather(1982)}]{mather_bolometer}
Mather, J.~C. 1982, Appl. Opt., 21, 1125

\bibitem[{{Miller} {et~al.}(2016){Miller}, {Chuss}, {Marriage}, {Wollack},
  {Appel}, {Bennett}, {Eimer}, {Essinger-Hileman}, {Fixsen}, {Harrington},
  {Moseley}, {Rostem}, {Switzer}, \& {Watts}}]{Miller2015}
{Miller}, N.~J., {Chuss}, D.~T., {Marriage}, T.~A., {et~al.} 2016, \apj, 818,
  151

\bibitem[{{Page} {et~al.}(2003{\natexlab{a}}){Page}, {Barnes}, {Hinshaw},
  {Spergel}, {Weiland}, {Wollack}, {Bennett}, {Halpern}, {Jarosik}, {Kogut},
  {Limon}, {Meyer}, {Tucker}, \& {Wright}}]{page_2003}
{Page}, L., {Barnes}, C., {Hinshaw}, G., {et~al.} 2003{\natexlab{a}}, \apjs,
  148, 39

\bibitem[{{Page} {et~al.}(2003{\natexlab{b}}){Page}, {Jackson}, {Barnes},
  {Bennett}, {Halpern}, {Hinshaw}, {Jarosik}, {Kogut}, {Limon}, {Meyer},
  {Spergel}, {Tucker}, {Wilkinson}, {Wollack}, \& {Wright}}]{page2003}
{Page}, L., {Jackson}, C., {Barnes}, C., {et~al.} 2003{\natexlab{b}}, \apj,
  585, 566

\bibitem[{Petroff {et~al.}(2019)Petroff, Appel, Rostem, Bennett, Eimer,
  Marriage, Ramirez, \& Wollack}]{petroff_bb}
Petroff, M., Appel, J., Rostem, K., {et~al.} 2019, Review of Scientific
  Instruments, 90, 024701

\bibitem[{{Planck Collaboration} {et~al.}(2015){Planck Collaboration}, {Ade},
  {Aghanim}, {Alina}, {Alves}, {Armitage-Caplan}, {Arnaud}, {Arzoumanian},
  {Ashdown}, {Atrio-Barandela}, \& et~al.}]{planck:IntermediateXIX}
{Planck Collaboration}, {Ade}, P.~A.~R., {Aghanim}, N., {et~al.} 2015, \aap,
  576, A104

\bibitem[{{Planck Collaboration} {et~al.}(2016{\natexlab{a}}){Planck
  Collaboration}, {Ade}, {Aghanim}, {Ashdown}, {Aumont}, {Baccigalupi},
  {Ballardini}, {Banday}, {Barreiro}, {Bartolo}, \&
  et~al.}]{planck_2015_results_II}
---. 2016{\natexlab{a}}, \aap, 594, A2

\bibitem[{{Planck Collaboration} {et~al.}(2016{\natexlab{b}}){Planck
  Collaboration}, {Ade}, {Aghanim}, {Arnaud}, {Arroja}, {Ashdown}, {Aumont},
  {Baccigalupi}, {Ballardini}, {Banday}, \& et~al.}]{planck2015XX}
---. 2016{\natexlab{b}}, \aap, 594, A20

\bibitem[{{Planck Collaboration} {et~al.}(2016{\natexlab{c}}){Planck
  Collaboration}, {Ade}, {Aghanim}, {Arg{\"u}eso}, {Arnaud}, {Ashdown},
  {Aumont}, {Baccigalupi}, {Banday}, {Barreiro}, \& et~al.}]{planck_tauA}
---. 2016{\natexlab{c}}, \aap, 594, A26

\bibitem[{{Planck Collaboration} {et~al.}(2018{\natexlab{a}}){Planck
  Collaboration}, {Akrami}, {Ashdown}, {Aumont}, {Baccigalupi}, {Ballardini},
  {Banday}, {Barreiro}, {Bartolo}, {Basak}, {Benabed}, {Bersanelli},
  {Bielewicz}, {Bond}, {Borrill}, {Bouchet}, {Boulanger}, {Bucher}, {Burigana},
  {Calabrese}, {Cardoso}, {Carron}, {Casaponsa}, {Challinor}, {Colombo},
  {Combet}, {Crill}, {Cuttaia}, {de Bernardis}, {de Rosa}, {de Zotti},
  {Delabrouille}, {Delouis}, {Di Valentino}, {Dickinson}, {Diego}, {Donzelli},
  {Dor{\'e}}, {Ducout}, {Dupac}, {Efstathiou}, {Elsner}, {En{\ss}lin},
  {Eriksen}, {Falgarone}, {Fernandez-Cobos}, {Finelli}, {Forastieri},
  {Frailis}, {Fraisse}, {Franceschi}, {Frolov}, {Galeotta}, {Galli}, {Ganga},
  {G{\'e}nova-Santos}, {Gerbino}, {Ghosh}, {Gonz{\'a}lez-Nuevo}, {G{\'o}rski},
  {Gratton}, {Gruppuso}, {Gudmundsson}, {Handley}, {Hansen}, {Helou},
  {Herranz}, {Huang}, {Jaffe}, {Karakci}, {Keih{\"a}nen}, {Keskitalo},
  {Kiiveri}, {Kim}, {Kisner}, {Krachmalnicoff}, {Kunz}, {Kurki-Suonio},
  {Lagache}, {Lamarre}, {Lasenby}, {Lattanzi}, {Lawrence}, {Le Jeune},
  {Levrier}, {Liguori}, {Lilje}, {Lindholm}, {L{\'o}pez-Caniego}, {Lubin},
  {Ma}, {Mac{\'{\i}}as-P{\'e}rez}, {Maggio}, {Maino}, {Mandolesi}, {Mangilli},
  {Marcos-Caballero}, {Martin}, {Mart{\'{\i}}nez-Gonz{\'a}lez}, {Matarrese},
  {Mauri}, {McEwen}, {Meinhold}, {Melchiorri}, {Mennella}, {Migliaccio},
  {Miville-Desch{\^e}nes}, {Molinari}, {Moneti}, {Montier}, {Morgante},
  {Natoli}, {Oppizzi}, {Pagano}, {Paoletti}, {Partridge}, {Peel}, {Pettorino},
  {Piacentini}, {Polenta}, {Puget}, {Rachen}, {Reinecke}, {Remazeilles},
  {Renzi}, {Rocha}, {Roudier}, {Rubi{\~n}o-Mart{\'{\i}}n}, {Ruiz-Granados},
  {Salvati}, {Sandri}, {Savelainen}, {Scott}, {Seljebotn}, {Sirignano},
  {Spencer}, {Suur-Uski}, {Tauber}, {Tavagnacco}, {Tenti}, {Thommesen},
  {Toffolatti}, {Tomasi}, {Trombetti}, {Valiviita}, {Van Tent}, {Vielva},
  {Villa}, {Vittorio}, {Wandelt}, {Wehus}, {Zacchei}, \&
  {Zonca}}]{planck_2018_diffuse}
{Planck Collaboration}, {Akrami}, Y., {Ashdown}, M., {et~al.}
  2018{\natexlab{a}}, ArXiv e-prints, arXiv:1807.06208

\bibitem[{{Planck Collaboration} {et~al.}(2018{\natexlab{b}}){Planck
  Collaboration}, {Aghanim}, {Akrami}, {Ashdown}, {Aumont}, {Baccigalupi},
  {Ballardini}, {Banday}, {Barreiro}, {Bartolo}, {Basak}, {Battye}, {Benabed},
  {Bernard}, {Bersanelli}, {Bielewicz}, {Bock}, {Bond}, {Borrill}, {Bouchet},
  {Boulanger}, {Bucher}, {Burigana}, {Butler}, {Calabrese}, {Cardoso},
  {Carron}, {Challinor}, {Chiang}, {Chluba}, {Colombo}, {Combet}, {Contreras},
  {Crill}, {Cuttaia}, {de Bernardis}, {de Zotti}, {Delabrouille}, {Delouis},
  {Di Valentino}, {Diego}, {Dor{\'e}}, {Douspis}, {Ducout}, {Dupac}, {Dusini},
  {Efstathiou}, {Elsner}, {En{\ss}lin}, {Eriksen}, {Fantaye}, {Farhang},
  {Fergusson}, {Fernandez-Cobos}, {Finelli}, {Forastieri}, {Frailis},
  {Franceschi}, {Frolov}, {Galeotta}, {Galli}, {Ganga}, {G{\'e}nova-Santos},
  {Gerbino}, {Ghosh}, {Gonz{\'a}lez-Nuevo}, {G{\'o}rski}, {Gratton},
  {Gruppuso}, {Gudmundsson}, {Hamann}, {Handley}, {Herranz}, {Hivon}, {Huang},
  {Jaffe}, {Jones}, {Karakci}, {Keih{\"a}nen}, {Keskitalo}, {Kiiveri}, {Kim},
  {Kisner}, {Knox}, {Krachmalnicoff}, {Kunz}, {Kurki-Suonio}, {Lagache},
  {Lamarre}, {Lasenby}, {Lattanzi}, {Lawrence}, {Le Jeune}, {Lemos},
  {Lesgourgues}, {Levrier}, {Lewis}, {Liguori}, {Lilje}, {Lilley}, {Lindholm},
  {L{\'o}pez-Caniego}, {Lubin}, {Ma}, {Mac{\'{\i}}as-P{\'e}rez}, {Maggio},
  {Maino}, {Mandolesi}, {Mangilli}, {Marcos-Caballero}, {Maris}, {Martin},
  {Martinelli}, {Mart{\'{\i}}nez-Gonz{\'a}lez}, {Matarrese}, {Mauri}, {McEwen},
  {Meinhold}, {Melchiorri}, {Mennella}, {Migliaccio}, {Millea}, {Mitra},
  {Miville-Desch{\^e}nes}, {Molinari}, {Montier}, {Morgante}, {Moss}, {Natoli},
  {N{\o}rgaard-Nielsen}, {Pagano}, {Paoletti}, {Partridge}, {Patanchon},
  {Peiris}, {Perrotta}, {Pettorino}, {Piacentini}, {Polastri}, {Polenta},
  {Puget}, {Rachen}, {Reinecke}, {Remazeilles}, {Renzi}, {Rocha}, {Rosset},
  {Roudier}, {Rubi{\~n}o-Mart{\'{\i}}n}, {Ruiz-Granados}, {Salvati}, {Sandri},
  {Savelainen}, {Scott}, {Shellard}, {Sirignano}, {Sirri}, {Spencer},
  {Sunyaev}, {Suur-Uski}, {Tauber}, {Tavagnacco}, {Tenti}, {Toffolatti},
  {Tomasi}, {Trombetti}, {Valenziano}, {Valiviita}, {Van Tent}, {Vibert},
  {Vielva}, {Villa}, {Vittorio}, {Wandelt}, {Wehus}, {White}, {White},
  {Zacchei}, \& {Zonca}}]{planck_2018_cosmo_param}
{Planck Collaboration}, {Aghanim}, N., {Akrami}, Y., {et~al.}
  2018{\natexlab{b}}, ArXiv e-prints, arXiv:1807.06209

\bibitem[{{Planck Collaboration} {et~al.}(2018{\natexlab{c}}){Planck
  Collaboration}, {Akrami}, {Arroja}, {Ashdown}, {Aumont}, {Baccigalupi},
  {Ballardini}, {Banday}, {Barreiro}, {Bartolo}, {Basak}, {Benabed}, {Bernard},
  {Bersanelli}, {Bielewicz}, {Bock}, {Bond}, {Borrill}, {Bouchet}, {Boulanger},
  {Bucher}, {Burigana}, {Butler}, {Calabrese}, {Cardoso}, {Carron},
  {Challinor}, {Chiang}, {Colombo}, {Combet}, {Contreras}, {Crill}, {Cuttaia},
  {de Bernardis}, {de Zotti}, {Delabrouille}, {Delouis}, {Di Valentino},
  {Diego}, {Donzelli}, {Dor{\'e}}, {Douspis}, {Ducout}, {Dupac}, {Dusini},
  {Efstathiou}, {Elsner}, {En{\ss}lin}, {Eriksen}, {Fantaye}, {Fergusson},
  {Fernandez-Cobos}, {Finelli}, {Forastieri}, {Frailis}, {Franceschi},
  {Frolov}, {Galeotta}, {Galli}, {Ganga}, {Gauthier}, {G{\'e}nova-Santos},
  {Gerbino}, {Ghosh}, {Gonz{\'a}lez-Nuevo}, {G{\'o}rski}, {Gratton},
  {Gruppuso}, {Gudmundsson}, {Hamann}, {Handley}, {Hansen}, {Herranz}, {Hivon},
  {Hooper}, {Huang}, {Jaffe}, {Jones}, {Keih{\"a}nen}, {Keskitalo}, {Kiiveri},
  {Kim}, {Kisner}, {Krachmalnicoff}, {Kunz}, {Kurki-Suonio}, {Lagache},
  {Lamarre}, {Lasenby}, {Lattanzi}, {Lawrence}, {Le Jeune}, {Lesgourgues},
  {Levrier}, {Lewis}, {Liguori}, {Lilje}, {Lindholm}, {Lpez-Caniego}, {Lubin},
  {Ma}, {Mac{\'{\i}}as-P{\'e}rez}, {Maggio}, {Maino}, {Mandolesi}, {Mangilli},
  {Marcos-Caballero}, {Maris}, {Martin}, {Mart{\'{\i}}nez-Gonz{\'a}lez},
  {Matarrese}, {Mauri}, {McEwen}, {Meerburg}, {Meinhold}, {Melchiorri},
  {Mennella}, {Migliaccio}, {Mitra}, {Miville-Desch{\^e}nes}, {Molinari},
  {Moneti}, {Montier}, {Morgante}, {Moss}, {M{\"u}nchmeyer}, {Natoli},
  {N{\o}rgaard-Nielsen}, {Pagano}, {Paoletti}, {Partridge}, {Patanchon},
  {Peiris}, {Perrotta}, {Pettorino}, {Piacentini}, {Polastri}, {Polenta},
  {Puget}, {Rachen}, {Reinecke}, {Remazeilles}, {Renzi}, {Rocha}, {Rosset},
  {Roudier}, {Rubi{\~n}o-Mart{\'{\i}}n}, {Ruiz-Granados}, {Salvati}, {Sandri},
  {Savelainen}, {Scott}, {Shellard}, {Shiraishi}, {Sirignano}, {Sirri},
  {Spencer}, {Sunyaev}, {Suur-Uski}, {Tauber}, {Tavagnacco}, {Tenti},
  {Toffolatti}, {Tomasi}, {Trombetti}, {Valiviita}, {Van Tent}, {Vielva},
  {Villa}, {Vittorio}, {Wandelt}, {Wehus}, {White}, {Zacchei}, {Zibin}, \&
  {Zonca}}]{planck_2018_inflation}
{Planck Collaboration}, {Akrami}, Y., {Arroja}, F., {et~al.}
  2018{\natexlab{c}}, ArXiv e-prints, arXiv:1807.06211

\bibitem[{{Planck Collaboration} {et~al.}(2018{\natexlab{d}}){Planck
  Collaboration}, {Akrami}, {Ashdown}, {Aumont}, {Baccigalupi}, {Ballardini},
  {Banday}, {Barreiro}, {Bartolo}, {Basak}, {Benabed}, {Bernard}, {Bersanelli},
  {Bielewicz}, {Bond}, {Borrill}, {Bouchet}, {Boulanger}, {Bracco}, {Bucher},
  {Burigana}, {Calabrese}, {Cardoso}, {Carron}, {Chiang}, {Combet}, {Crill},
  {de Bernardis}, {de Zotti}, {Delabrouille}, {Delouis}, {Di Valentino},
  {Dickinson}, {Diego}, {Ducout}, {Dupac}, {Efstathiou}, {Elsner},
  {En{\ss}lin}, {Falgarone}, {Fantaye}, {Ferri{\`e}re}, {Finelli},
  {Forastieri}, {Frailis}, {Fraisse}, {Franceschi}, {Frolov}, {Galeotta},
  {Galli}, {Ganga}, {G{\'e}nova-Santos}, {Ghosh}, {Gonz{\'a}lez-Nuevo},
  {G{\'o}rski}, {Gruppuso}, {Gudmundsson}, {Guillet}, {Handley}, {Hansen},
  {Herranz}, {Huang}, {Jaffe}, {Jones}, {Keih{\"a}nen}, {Keskitalo}, {Kiiveri},
  {Kim}, {Krachmalnicoff}, {Kunz}, {Kurki-Suonio}, {Lamarre}, {Lasenby}, {Le
  Jeune}, {Levrier}, {Liguori}, {Lilje}, {Lindholm}, {L{\'o}pez-Caniego},
  {Lubin}, {Ma}, {Mac{\'{\i}}as-P{\'e}rez}, {Maggio}, {Maino}, {Mandolesi},
  {Mangilli}, {Martin}, {Mart{\'{\i}}nez-Gonz{\'a}lez}, {Matarrese}, {McEwen},
  {Meinhold}, {Melchiorri}, {Migliaccio}, {Miville-Desch{\^e}nes}, {Molinari},
  {Moneti}, {Montier}, {Morgante}, {Natoli}, {Pagano}, {Paoletti}, {Pettorino},
  {Piacentini}, {Polenta}, {Puget}, {Rachen}, {Reinecke}, {Remazeilles},
  {Renzi}, {Rocha}, {Rosset}, {Roudier}, {Rubi{\~n}o-Mart{\'{\i}}n},
  {Ruiz-Granados}, {Salvati}, {Sandri}, {Savelainen}, {Scott}, {Soler},
  {Spencer}, {Tauber}, {Tavagnacco}, {Toffolatti}, {Tomasi}, {Trombetti},
  {Valiviita}, {Vansyngel}, {Van Tent}, {Vielva}, {Villa}, {Vittorio}, {Wehus},
  {Zacchei}, \& {Zonca}}]{planck_2018_dust}
{Planck Collaboration}, {Akrami}, Y., {Ashdown}, M., {et~al.}
  2018{\natexlab{d}}, ArXiv e-prints, arXiv:1801.04945

\bibitem[{{Planck Collaboration Int.~L}(2017)}]{planckL2016}
{Planck Collaboration Int.~L}. 2017, \aap, 599, A51

\bibitem[{{Polarbear Collaboration} {et~al.}(2014){Polarbear Collaboration},
  {Ade}, {Akiba}, {Anthony}, {Arnold}, {Atlas}, {Barron}, {Boettger},
  {Borrill}, {Chapman}, {Chinone}, {Dobbs}, {Elleflot}, {Errard}, {Fabbian},
  {Feng}, {Flanigan}, {Gilbert}, {Grainger}, {Halverson}, {Hasegawa},
  {Hattori}, {Hazumi}, {Holzapfel}, {Hori}, {Howard}, {Hyland}, {Inoue},
  {Jaehnig}, {Jaffe}, {Keating}, {Kermish}, {Keskitalo}, {Kisner}, {Le Jeune},
  {Lee}, {Leitch}, {Linder}, {Lungu}, {Matsuda}, {Matsumura}, {Meng}, {Miller},
  {Morii}, {Moyerman}, {Myers}, {Navaroli}, {Nishino}, {Orlando}, {Paar},
  {Peloton}, {Poletti}, {Quealy}, {Rebeiz}, {Reichardt}, {Richards}, {Ross},
  {Schanning}, {Schenck}, {Sherwin}, {Shimizu}, {Shimmin}, {Shimon},
  {Siritanasak}, {Smecher}, {Spieler}, {Stebor}, {Steinbach}, {Stompor},
  {Suzuki}, {Takakura}, {Tomaru}, {Wilson}, {Yadav}, \&
  {Zahn}}]{polarbear_2014}
{Polarbear Collaboration}, {Ade}, P.~A.~R., {Akiba}, Y., {et~al.} 2014, \apj,
  794, 171

\bibitem[{{Rees}(1968)}]{rees_cmbpol}
{Rees}, M.~J. 1968, \apjl, 153, L1

\bibitem[{{Richards}(1994)}]{richards_bolometer}
{Richards}, P.~L. 1994, Journal of Applied Physics, 76, 1

\bibitem[{{Ritacco} {et~al.}(2018){Ritacco}, {Mac{\'{\i}}as-P{\'e}rez},
  {Ponthieu}, {Adam}, {Ade}, {Andr{\'e}}, {Aumont}, {Beelen}, {Beno{\^i}t},
  {Bideaud}, {Billot}, {Bourrion}, {Bracco}, {Calvo}, {Catalano}, {Coiffard},
  {Comis}, {D'Addabbo}, {De Petris}, {D{\'e}sert}, {Doyle}, {Goupy}, {Kramer},
  {Lagache}, {Leclercq}, {Lestrade}, {Mauskopf}, {Mayet}, {Maury},
  {Monfardini}, {Pajot}, {Pascale}, {Perotto}, {Pisano}, {Rebolo-Iglesias},
  {Rev{\'e}ret}, {Rodriguez}, {Romero}, {Roussel}, {Ruppin}, {Schuster},
  {Sievers}, {Siringo}, {Thum}, {Triqueneaux}, {Tucker}, {Wiesemeyer}, \&
  {Zylka}}]{nika_tauA}
{Ritacco}, A., {Mac{\'{\i}}as-P{\'e}rez}, J.~F., {Ponthieu}, N., {et~al.} 2018,
  \aap, 616, A35

\bibitem[{{Rostem} {et~al.}(2012){Rostem}, {Bennett}, {Chuss}, {Costen},
  {Crowe}, {Denis}, {Eimer}, {Lourie}, {Essinger-Hileman}, {Marriage},
  {Moseley}, {Stevenson}, {Towner}, {Voellmer}, {Wollack}, \&
  {Zeng}}]{karwan_spie}
{Rostem}, K., {Bennett}, C.~L., {Chuss}, D.~T., {et~al.} 2012, in Society of
  Photo-Optical Instrumentation Engineers (SPIE) Conference Series, Vol. 8452,
  Society of Photo-Optical Instrumentation Engineers (SPIE) Conference Series

\bibitem[{{Rostem} {et~al.}(2014{\natexlab{a}}){Rostem}, {Chuss}, {Colazo},
  {Crowe}, {Denis}, {Lourie}, {Moseley}, {Stevenson}, \&
  {Wollack}}]{leg_precision}
{Rostem}, K., {Chuss}, D.~T., {Colazo}, F.~A., {et~al.} 2014{\natexlab{a}},
  Journal of Applied Physics, 115, 124508

\bibitem[{{Rostem} {et~al.}(2014{\natexlab{b}}){Rostem}, {Ali}, {Appel},
  {Bennett}, {Chuss}, {Colazo}, {Crowe}, {Denis}, {Essinger-Hileman},
  {Marriage}, {Moseley}, {Stevenson}, {Towner}, {U-Yen}, \&
  {Wollack}}]{karwan_spie_2014}
{Rostem}, K., {Ali}, A., {Appel}, J.~W., {et~al.} 2014{\natexlab{b}}, in
  \procspie, Vol. 9153, Millimeter, Submillimeter, and Far-Infrared Detectors
  and Instrumentation for Astronomy VII, 91530B

\bibitem[{{Sato}(1981)}]{sato:1981}
{Sato}, K. 1981, \mnras, 195, 467

\bibitem[{{Starobinsky}(1982)}]{starobinsky:1982}
{Starobinsky}, A.~A. 1982, Physics Letters B, 117, 175

\bibitem[{{Troitskii}(1967)}]{troitskii_moon_model}
{Troitskii}, V.~S. 1967, Radiophysics and Quantum Electronics, 10, 709

\bibitem[{{Troitskii} \& {Tikhonova}(1970)}]{Troitsky_1970}
{Troitskii}, V.~S., \& {Tikhonova}, T.~V. 1970, Radiophysics and Quantum
  Electronics, 13, 981

\bibitem[{{Troitsky} {et~al.}(1968){Troitsky}, {Burov}, \&
  {Alyoshina}}]{Troitsky_1968}
{Troitsky}, V.~S., {Burov}, A.~B., \& {Alyoshina}, T.~N. 1968, \icarus, 8, 423

\bibitem[{{Tsang} {et~al.}(2016){Tsang}, {Hu}, \& {Zheng}}]{changE_calib_2}
{Tsang}, K., {Hu}, G.-P., \& {Zheng}, Y.-C. 2016, in AAS/Division for Planetary
  Sciences Meeting Abstracts, Vol.~48, AAS/Division for Planetary Sciences
  Meeting Abstracts \#48, 223.06

\bibitem[{{van Vliet}(1967)}]{vanvliet}
{van Vliet}, K.~M. 1967, \ao, 6, 1145

\bibitem[{{Watts} {et~al.}(2015){Watts}, {Larson}, {Marriage}, {Abitbol},
  {Appel}, {Bennett}, {Chuss}, {Eimer}, {Essinger-Hileman}, {Miller}, {Rostem},
  \& {Wollack}}]{Watts2015}
{Watts}, D.~J., {Larson}, D., {Marriage}, T.~A., {et~al.} 2015, \apj, 814, 103

\bibitem[{Watts {et~al.}(2018)Watts, Wang, Ali, Appel, Bennett, Chuss, Dahal,
  Eimer, Essinger-Hileman, Harrington, Hinshaw, Iuliano, Marriage, Miller,
  Padilla, Parker, Petroff, Rostem, Wollack, \& Xu}]{Watts2018}
Watts, D.~J., Wang, B., Ali, A., {et~al.} 2018, \apj, 863, 121

\bibitem[{{Weiland} {et~al.}(2011){Weiland}, {Odegard}, {Hill}, {Wollack},
  {Hinshaw}, {Greason}, {Jarosik}, {Page}, {Bennett}, {Dunkley}, {Gold},
  {Halpern}, {Kogut}, {Komatsu}, {Larson}, {Limon}, {Meyer}, {Nolta}, {Smith},
  {Spergel}, {Tucker}, \& {Wright}}]{wmap_weiland}
{Weiland}, J.~L., {Odegard}, N., {Hill}, R.~S., {et~al.} 2011, \apjs, 192, 19

\bibitem[{{Zaldarriaga} \& {Seljak}(1997)}]{seljak_zaldariaga}
{Zaldarriaga}, M., \& {Seljak}, U. 1997, \prd, 55, 1830

\bibitem[{{Zeng}(2012)}]{zeng_phd}
{Zeng}, L. 2012, PhD thesis, The Johns Hopkins University

\bibitem[{{Zheng} {et~al.}(2012){Zheng}, {Tsang}, {Chan}, {Zou}, {Zhang}, \&
  {Ouyang}}]{changE}
{Zheng}, Y.~C., {Tsang}, K.~T., {Chan}, K.~L., {et~al.} 2012, \icarus, 219, 194

\bibitem[{{Zmuidzinas}(2003)}]{zmuidzinas}
{Zmuidzinas}, J. 2003, \ao, 42, 4989

\end{thebibliography}
\bibliographystyle{aasjournal}

\end{document}